\documentclass[twocolumn,amsmath,amssymb,
aps, showpacs,graphicx]{revtex4-2}

\usepackage{graphicx}
\usepackage{dcolumn}
\usepackage{bm}
\usepackage{lineno, hyperref}
\usepackage{natbib}
\begin{document}

\title{Excitonic condensation and metal-semiconductor transition \\ in AA bilayer graphene in the external magnetic field}
\author{V. Apinyan} 
\altaffiliation[e-mail:]{v.apinyan@intibs.pl}
\author{T. K. Kope\'{c}} 
\affiliation{Institute of Low Temperature and Structure Research, Polish Academy of Sciences, \\ 50-422 Wroc\l{}aw, ul. Ok\'{o}lna 2, Poland}

\begin{abstract}
In this paper, the effects of the external transverse magnetic field $B$ (perpendicular to the surface of the layers) on the electronic and excitonic properties are studied in the AA-stacked bilayer graphene (BLG). The effects of the Coulomb interactions and excitonic pairing have been taken into account and analyzed in detail within the bilayer Hubbard model. Both half-filling and partial filling regimes have been taken into account and the magnetic field dependence of a series of physical parameters was found. It is shown that the difference between the average electron concentrations in the layers vanishes at some critical value of magnetic field $B_{c}$ and the chemical potential is calculated numerically above and below that value. The role of the Coulomb interactions on the average carrier concentrations in the layers has been analyzed, and the excitonic order parameters have been calculated for different spin orientations. We found a possibility for the particle population inversion between the layers when varying the external magnetic field. The calculated electronic band structure in the AA-BLG shows the presence of metal-semiconductor transition, governed by the strength of the applied magnetic field or the interlayer interaction potential. We show that for high magnetic fields the band-gap is approaching the typical values of the gaps in the usual semiconductors. It is demonstrated that, at some parameter regimes, the AA-BLG behaves like a spin-valve device, by permitting the electron transport with only one spin direction. All calculations have been performed at the zero-temperature limit.  
\end{abstract}



\maketitle

\section{\label{sec:Section_1} Introduction}
%
The bilayer graphene (BLG) systems have been the subject for many theoretical and experimental investigations, due to their interesting physical properties \cite{cite_1, cite_2, cite_3, cite_4, cite_5, cite_6, cite_7, cite_8}. 
Recently, stable AA-BLG structures have been obtained experimentally, which are  promising for technological applications
\cite{cite_9, cite_10, cite_11, cite_12}.  
The opening of the band-gap in the single-particle excitation spectrum of AB-stacked BLG systems under the influence of the electric field classified the AB-BLG among the  semiconducting materials with the controllable gap parameter, which opened new opportunities for their technological applications \cite{cite_1, cite_2, cite_3, cite_4}. Despite many efforts to observe such a gap in the single-particle excitation spectrum of AA-BLG, the band structure of this system has been found with no excitation gap \cite{cite_13} and with linear energy spectrum. Recently a possible band-gap opening issue is reported in \cite{cite_14} where the authors evaluate the transmission and reflection probabilities in the AA-BLG with the layers encapsulated in the dielectric. The band-gap found in their research is about $40$ meV and is caused by the induced mass-terms via the dielectric medium. This value of the gap is slightly larger than the value of the band-gap found in single-layer graphene when using the substrates of SiC or h-BN ($E_{g}\sim 20$ meV) \cite{cite_15, cite_16}. Another interesting band-gap opening is related to the effect of spin-orbit coupling in graphene which is indeed found to be very small ($E_{\rm SO}\sim10 \mu$eV)\cite{cite_17, cite_18}. In a series of works the authors show the opening of the very small band-gaps when including the effects of the spin-orbit coupling (SOC) \cite{cite_17, cite_18, cite_19, cite_20}. These results are promising, mostly, for studying the spin-Hall states in AA-BLG systems \cite{cite_19}. 

Concerning the theoretical side of studies, many interesting works have been done to study the metal-insulator transition and band-gap opening (in the low-$U$ limit) in doped AA-BLG \cite{cite_21, cite_22}, the influence of antiferromagnetism on the physical properties in AA-BLG \cite{cite_23, cite_24, cite_25, cite_26}, the coexistence of the antiferromagnetic and excitonic insulator phases \cite{cite_23, cite_24, cite_25}, the dynamical instabilities in the AA-BLG \cite{cite_25}, the optical plasmonic gap opening \cite{cite_27}, etc. Recently, an electronic transport gap has been demonstrated in \cite{cite_28}, and the charge-carrier dependence on the spin relaxation time was analyzed. Moreover, the bilayer-graphene based spin-valve devices have been proposed in this context. In the work \cite{cite_29} a very large band-gap opening has been shown in unbiased AA-bilayer graphene with variable interlayer separation distance $c_0$, starting from van der Waals-like distances with $c_0=3.58\AA$ to small separation distances, allowing the chemical bondings between the carbon atoms situated in different layers in the bilayer. A series of phase transitions have been found, from the semi-metal to the wide band-gap semiconducting one, when passing from van der Waals to covalent bond regimes.  

The effect of the magnetic field on the electronic and transport properties in the AA-BLG has been studied in a series of theoretical works \cite{cite_30, cite_31, cite_32, cite_33, cite_34, cite_35, cite_36, cite_37, cite_38, cite_39, cite_40, cite_41}. The magnetic filed dependence of thermal properties in the doped AA-BLG has been considered theoretically in \cite{cite_31, cite_35}, where the presence of both electron and hole type contributions in transport properties has been shown in AA-BLG. However, up to now, little work pays attention to the influence of the magnetic field on the charge redistribution and band-gap opening in the pristine AA-BLG. 

In this work, we consider the effects of the magnetic field on the excitonic properties in the system and we show the possibilities of the excitonic condensates states in AA-BLG at different values of the magnetic field. We show the possibility of formation of different direct band-gaps in the energy spectrum of the AA-BLG under the influence of the transverse magnetic field (see in Fig.~\ref{fig:Fig_1}). First of all we show the existence of a critical value of the magnetic field above which the physical parameters in the system change their behavior. We show theoretically that the complete charge neutrality (CN) (when the average total charge densities in both layers get equal values) could be achieved at the half-filling case and the average electron populations inversion occurs only in the case of partial-filling. The CN in the AA-BLG system could be also obtained when varying the interlayer interaction potential. This effect is also discussed in the present paper. The CN of the entire bilayer graphene system is important for its non-destructive applications in modern technological devices. To achieve the CN in such construction a huge number of internal and external physical parameters should be properly considered and tuned \cite{cite_37}. This requires also an important experimental setup and efforts to calibrate the BLG system at the CN \cite{cite_38}. The electronic and transport properties at the CN and away of it have been discussed in a number of works \cite{cite_42, cite_43, cite_44, cite_45, cite_46, cite_47}. 

Furthermore, we calculate the band-gaps in the system and we show that they are spin-dependent and totally controllable by the strength of the applied magnetic field. Our numericalcalculations demonstrate that the opened band-gaps are comparable with the band-gaps in the typical direct low-band-gap semiconducting systems of types A$^{\rm III}$B$^{\rm V}$ or A$^{\rm IV}$B$^{\rm VI}$ \cite{cite_48}.
Particularly, we show that in the regime of non-interacting layers (and in the presence of the magnetic field), a very large band-gap is opening in the electronic band structure with the value of the band-gap $E_{g}\sim 200$ meV (much larger than those obtained from the SOC considerations \cite{cite_17, cite_18, cite_19, cite_20}), The transition from semiconducting to weak-metallic states occurs when increasing the interlayer Coulomb potential. Meanwhile, when varying the magnetic field, from zero up to high values, an opposite transition occurs (at the half-filling), from metallic to the semiconducting state with a sufficiently large band-gap of order $E_{g}\sim150.6$ meV, typical for the semiconducting systems of types A$^{\rm III}$B$^{\rm V}$ or A$^{\rm IV}$B$^{\rm VI}$ \cite{cite_48}. Our results suggest that, for the given range of parameters and conditions imposed on the system, one can quench the electronic conductivity in one spin channel (for example $\sigma=\uparrow$), while the other one ($\sigma=\downarrow$) remains open. Those results are purposeful for the use of AA-BLG as the spin-valve device \cite{cite_49, cite_50}, for examining the spin-controlled quantum transport at the nanoscale \cite{cite_51, cite_52, cite_53, cite_54, cite_55, cite_56} and excitonic condensation phenomena, not observed yet experimentally \cite{cite_57, cite_58, cite_59}.        

The present paper is organized as follows: in Section \ref{sec:Section_2} we introduce the AA bilayer graphene within the generalized Hubbard model. In Section \ref{sec:Section_3},  we obtain the electronic band structure and we give the set of self-consistent equations. In Section \ref{sec:Section_4}, we present the numerical results for the important physical quantities in the system and we discuss different band-gap formations. In Section \ref{sec:Section_5}, we give a conclusion to our paper. In Appendix \ref{sec:Section_6}, we obtain the system of self-consistent equations and in Appendix \ref{sec:Section_7} the mean-field decouplings procedures are discussed. 
%
\section{\label{sec:Section_2} Hamiltonian of the interacting AA-BLG}
%
\subsection{\label{sec:Section_2_1} Bilayer Hubbard Hamiltonian with magnetic field}
\begin{figure}
	\begin{center}
		\includegraphics[width=8.8cm, height=4.8cm]{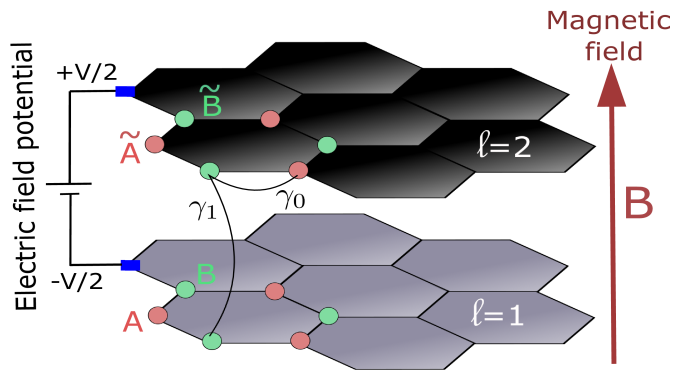}
		\caption{\label{fig:Fig_1}(Color online) The structure of biased AA-stacked bilayer graphene system with the applied external electric field potential $V$ (with potential $+V/2$ applied to the layer $\ell=2$ and $-V/2$ at the bottom layer $\ell=1$) and transverse magnetic field $B$ (see thick red-arrow in the picture), in the direction of the $z$-axis. Different layers in the AA-BLG have been shown with two sublattice sites $A$, $B$ in the layer $\ell=1$ and $\tilde{A}$, $\tilde{B}$ in the layer $\ell=2$.}
	\end{center}
\end{figure}

%
We consider here the bilayer graphene structure composed of two layers, and each atom in the top layer is stacked on top of the similar atom in the bottom layer. Moreover, there is no shift in their local atomic space position ${\bf{r}}$. This type of stacking order is called as the AA stacking.  
The Hamiltonian of the AA-bilayer graphene in the presence of the external static electric field-potential $V$ and magnetic field ${\bf{B}}$ (oriented along the $z$-axis, perpendicular to the planes of the layers) could be written in the following form 
%
%
\begin{eqnarray}
&&\hat{\cal{H}}_{{\rm AA}}=\hat{\cal{H}}_{\gamma_0}+\hat{\cal{H}}_{\gamma_1}+{\hat{\cal{H}}}_{U}+{\hat{\cal{H}}}_{W}+\hat{\cal{H}}_{V}+{\hat{\cal{H}}}_{\rm int}+\hat{\cal{H}}_{\mu},
	\label{Equation_1}
	\nonumber\\
\end{eqnarray} 
where 
\begin{eqnarray}
\hat{\cal{H}}_{\gamma_0}=&&-\gamma_0\sum_{\left\langle {\bf{r}},{\bf{r}}'\right\rangle}\sum_{\sigma}\left(\hat{a}^{\dag}_{\sigma}({\bf{r}})\hat{b}_{\sigma}({\bf{r}}')+h.c.\right)
\nonumber\\
&&-\gamma_0\sum_{\left\langle {\bf{r}},{\bf{r}}'\right\rangle}\sum_{\sigma}\left(\hat{\tilde{a}}^{\dag}_{\sigma}({\bf{r}})\hat{\tilde{b}}_{\sigma}({\bf{r}}')+h.c.\right)
	\label{Equation_2}
\end{eqnarray}
is the intralayer electron hopping Hamiltonian and the interlayer hopping term is given by the Hamiltonian 
\begin{eqnarray}
	&&\hat{\cal{H}}_{\gamma_1}=-\gamma_{1}\sum_{{\bf{r}}\sigma}\left(\hat{a}^{\dag}_{\sigma}({\bf{r}})\hat{\tilde{a}}_{\sigma}({\bf{r}})+\hat{b}^{\dag}_{\sigma}({\bf{r}})\hat{\tilde{b}}_{\sigma}({\bf{r}})+h.c.\right).
		\label{Equation_3}
\end{eqnarray}
The operators $\hat{a}_{\sigma}({\bf{r}}),\hat{b}_{\sigma}({\bf{r}}),\hat{\tilde{a}}_{\sigma}({\bf{r}})$ and $\hat{\tilde{b}}_{\sigma}({\bf{r}})$ describe the destruction of electrons at the given lattice site positions and the operators $\hat{a}^{\dag}_{\sigma}({\bf{r}}),\hat{b}^{\dag}_{\sigma}({\bf{r}}),\hat{\tilde{a}}^{\dag}_{\sigma}({\bf{r}})$, $\hat{\tilde{b}}^{\dag}_{\sigma}({\bf{r}})$ are the electron creation operators. The index $\sigma$ denotes the spin-variable, which takes two possible directions: $\sigma=\uparrow$ or $\sigma=\downarrow$. The electrons enter into covalent bonds in the graphene's layers and are attached with the atoms near atomic sites $A$, $B$ (in the layer with $\ell=1$) and $\tilde{A}$, $\tilde{B}$ (in the layer with $\ell=2$) (see the picture, in Fig.~\ref{fig:Fig_1}). 
The parameter $\gamma_0$ is the energy necessary for the hopping of electrons between the adjacent lattice sites in the layers. The summations $\left\langle ...\right\rangle$ in Eq.(\ref{Equation_2}) are taken over the nearest neighbor lattice sites positions ${\bf{r}}$, ${\bf{r}}'$ in the separate graphene layers. The energy parameter $\gamma_1$, in Eq.(\ref{Equation_3}), describes the local hopping of electrons between adjacent layers $\ell=1$ and $\ell=2$. The values, found experimentally, for the hopping parameters are $\gamma_0\sim 3$ eV and $\gamma_1=0.257$ eV, as it was reported in \cite{cite_60}.
Next, the summations in Eq.(\ref{Equation_3}) are over the lattice sites positions ${\bf{r}}=1,2...,N$ and the electron spin configurations with $\sigma=\uparrow,\downarrow$.
The Coulomb interaction terms ${\hat{\cal{H}}}_{U}$ and ${\hat{\cal{H}}}_{W}$ are give as
\begin{equation}
{\hat{\cal{H}}}_{U}=U\sum_{{\bf{r}}\eta}\hat{n}_{\eta\uparrow}({\bf{r}})\hat{n}_{\eta\downarrow}({\bf{r}}),
\label{Equation_4}
\end{equation}
and
\begin{equation}
{\hat{\cal{H}}}_{W}={W}\sum_{{\bf{r}}\sigma\sigma'}\hat{n}_{a\sigma}\left({\bf{r}}\right)\hat{n}_{\tilde{a}\sigma'}\left({\bf{r}}\right)+{W}\sum_{{\bf{r}}\sigma\sigma'}\hat{n}_{b\sigma}\left({\bf{r}}\right)\hat{n}_{\tilde{b}\sigma'}\left({\bf{r}}\right),
\label{Equation_5}
\end{equation}
where $U$ in Eq.(\ref{Equation_4}) is the on-site Coulomb repulsion between the electrons in the layers $\ell=1,2$. We suppose that the value of it is uniform in both layers. The second summation in Eq.(\ref{Equation_4}) is over all sublattice variables $\eta=a,b,\tilde{a}, \tilde{b}$. The parameter $W$, in Eq.(\ref{Equation_5}) denotes the local interlayer Coulomb interaction between the electrons at the same sublattice sites $A$,$\tilde{A}$ and $B$, $\tilde{B}$ in different layers.
The operators $\hat{n}_{\eta\uparrow}\left({\bf{r}}\right)$ and $\hat{n}_{\eta\downarrow}\left({\bf{r}}\right)$ in Eq.(\ref{Equation_4}) are the electron density operators for $\sigma=\uparrow$ and $\sigma=\downarrow$ and they are given as 
\begin{eqnarray}
	\hat{n}_{\eta\sigma}({\bf{r}})={\hat{\eta}}^{\dag}_{\sigma}({\bf{r}}){\hat{\eta}}_{\sigma}({\bf{r}}).
	\label{Equation_6}
\end{eqnarray}
The Coulomb potential $W$, in Eq.(\ref{Equation_5}), is coupled to the product of particle density operators $\hat{n}_{\eta\sigma}$ and $\hat{n}_{\tilde{\eta}\sigma'}$ in different layers (with $\eta\neq\tilde{\eta}$ and $\sigma,\sigma'=\uparrow\downarrow$), defined in Eq.(\ref{Equation_6}). 

Next, the couplings with the external electric potential and magnetic fields are given by the terms $\hat{\cal{H}}_{V}$ and $\hat{\cal{H}}_{\rm int}$ in Eq.(\ref{Equation_1}). We have
\begin{eqnarray}
\hat{\cal{H}}_{V}=\frac{V}{2}\sum_{{\bf{r}}}\left(\hat{n}_{2}\left({\bf{r}}\right)-\hat{n}_{1}\left({\bf{r}}\right)\right),
	\label{Equation_7}
\end{eqnarray}
where $V$ is the electric field potential with the value $+V/2$ at the top layer $\ell=2$ and the value $-V/2$ at the bottom layer $\ell=1$ (see in Fig.~\ref{fig:Fig_1}). The operators $\hat{n}_{1}\left({\bf{r}}\right)$ and $\hat{n}_{2}\left({\bf{r}}\right)$ in Eq.(\ref{Equation_7}) are total density operators in the individual layers, i.e.,
\begin{eqnarray}
	\hat{n}_{1}\left({\bf{r}}\right)=\hat{n}_{a}\left({\bf{r}}\right)+\hat{n}_{b}\left({\bf{r}}\right),
	\nonumber\\ \hat{n}_{2}\left({\bf{r}}\right)=\hat{n}_{\tilde{a}}\left({\bf{r}}\right)+\hat{n}_{\tilde{b}}\left({\bf{r}}\right).
	\label{Equation_8}
\end{eqnarray}
In turn, the operators $\hat{n}_{\eta}\left({\bf{r}}\right)$, in the right-hand sides in Eq.(\ref{Equation_8}), are given in the form 
\begin{eqnarray}
	\hat{n}_{\eta}\left({\bf{r}}\right)=\sum_{\sigma}\hat{n}_{\eta\sigma}({\bf{r}}).
	\label{Equation_9}
\end{eqnarray}
Furthermore, the coupling with the magnetic field is given as 
\begin{eqnarray}
	\hat{{\cal{H}}}_{\rm int}=-g\mu_{B}B_z\sum_{{\bf{r}}\eta}\left(\hat{n}_{\eta\uparrow}-\hat{n}_{\eta\downarrow}\right). 
	\label{Equation_10}
\end{eqnarray}
The parameter $g$ in Eq.(\ref{Equation_10}) is the Land\'{e} g-factor and it's value can be derived naturally from Dirac's equation ($g\sim2$, according with recent experimental measurements of this physical parameter in graphene structures \cite{cite_61}). Next, $\mu_{B}$, in Eq.(\ref{Equation_10}), is the Bohr magneton (intrinsic magnetic moment of an electron), which we put equal to $1$ (here, we use the convention $\hbar=1$), throughout this paper.
We considered here the magnetic field in the $z$ direction, perpendicular to the layers in the AA-BLG (see the thick-red arrow in Fig.~\ref{fig:Fig_1})
The Hamiltonian $\hat{\cal{H}}_{\mu}$, in Eq.(\ref{Equation_1}), is the chemical potential term and it is given as
\begin{eqnarray}
	\hat{{H}}_{\mu}=-\mu\sum_{{\bf{r}}}\hat{n}\left({\bf{r}}\right).
	\label{Equation_11}
\end{eqnarray}	
%
%

We ignored in Eq.(\ref{Equation_1}) the spin-obit coupling due to the fact that the effect on the band-gap, resulting from this type of interaction, is very small \cite{cite_17, cite_18, cite_19}.	

Furthermore, we introduce the fermionic Grassmann complex variables at the place of usual electron operators and we write the expression of the partition function of the system, in terms of the Grassmann field variables (see, a similar description, in \cite{cite_63}). We write the partition function ${\cal{Z}}$ of the AA-BLG system in the formalism based on the fermionic path integrals (see, in Ref.\cite{cite_63, cite_64})  
\begin{eqnarray}
{\cal{Z}}=\mathrm{Tr}{e^{-\beta{\cal{H}}}}=\int\prod_{\eta}\left[{\cal D}\bar{\eta}{\cal{D}}\eta\right] e^{-{\cal{S}}},
	\label{Equation_12}
\end{eqnarray}
where ${\cal{H}}$ is the total Hamiltonian of the AA-BLG and parameter $\beta$, in the exponential, in Eq.(\ref{Equation_12}), is given after the imaginary-time Matsubara formalism (with $0<\tau<\beta$) \cite{cite_66}, and we have $\beta=1/T$, where and $T$ is the thermodynamic temperature. Here, we used the units where the Boltzmann is $k_{B}=1$.

Furthermore, we note ${\cal{S}}={\cal{S}}\left[\bar{a},a,{\bar{b}},b,\bar{\tilde{a}},\tilde{a},\bar{\tilde{b}},\tilde{b}\right]$ as the total fermionic action of the system in terms of Grassmann variables. It is given in the imaginary time Matsubara representation. It can be expressed with the help of total Hamiltonian ${\cal{H}}_{{\cal{AA}}}(\tau)$ as
\begin{eqnarray}
	{\cal{S}}=\int^{\beta}_{0}d\tau{{\cal{H}}_{{\cal{AA}}}(\tau)}+\sum_{\eta}{\cal{S}}_{\rm B}\left[\bar{\eta},\eta\right].
	\label{Equation_13}
\end{eqnarray}
Next, ${\cal{S}}_{\rm B}\left[\bar{\eta},\eta\right]$, are the electronic Berry terms \cite{cite_64} and they are given by 
\begin{eqnarray}
	{\cal{S}}_{\rm B}\left[\bar{\eta},\eta\right]=\sum_{{\bf{r}}\sigma}\int^{\beta}_{0}d\tau \bar{\eta}_{\sigma}({\bf{r}}\tau)\partial_{\tau}\eta_{\sigma}({\bf{r}}\tau).
	\label{Equation_14}
\end{eqnarray}
%
%
\subsection{\label{sec:Section_2_2} Linearization of the quadratic density terms}
%
In fact, the nonlinear density terms, figuring in the Hubbard interaction Hamiltonians, in Eqs.(\ref{Equation_4}) and (\ref{Equation_5}) could be decoupled with the help of a series of Hubbard-Stratanovich transformations, which are indeed equivalent to the usual mean-field approximations. A detailed description of such procedures in given in the Appendix \ref{sec:Section_7}. Here, we present only the resulting contributions to the total Hamiltonian and action of the BLG system, coming from such decouplings. First, the contribution, from the linearization procedure of the Coulomb-$U$ term in Eq.(\ref{Equation_4}) is following
  \begin{eqnarray}
  	\delta{\cal{S}}_{U}=-\sum_{{\bf{r}}\eta}\int^{\beta}_{0}d\tau\frac{U}{2}\bar{n}_{\eta}n_{\eta}\left({\bf{r}}\tau\right)
  	\label{Equation_15}
  \end{eqnarray}
  and the contribution to total Hamiltonian in Eq.(\ref{Equation_1}) is straightforward
  \begin{eqnarray}
  	\delta{{\cal{H}}}_{U}=\frac{U}{2}\sum_{{\bf{r}}\eta}\bar{n}_{\eta}n_{\eta}\left({\bf{r}}\tau\right).
  	\label{Equation_16}
  \end{eqnarray} 
Here, the notation $\bar{n}_{\eta}$ means the grand canonical average of the density function ${n}_{\eta}\left({\bf{r}}\tau\right)$, i.e., in usual writing, we have $\bar{n}_{\eta}=\left\langle{n}_{\eta}\right\rangle$ . It could be calculated exactly with the help of the partition function, in Eq.(\ref{Equation_12}), as 
\begin{eqnarray}
	\left\langle ... \right\rangle=\frac{1}{{\cal{Z}}}\int\prod_{\eta}\left[{\cal D}\bar{\eta}{\cal{D}}\eta\right]...e^{-{\cal{S}}}.
	\label{Equation_17}
\end{eqnarray} 
When deriving the result in Eq.(\ref{Equation_16}), we have supposed that the average densities, corresponding to opposite spin orientations are equal $\left\langle {n}_{\eta \uparrow}\left({\bf{r}}\tau\right)\right\rangle=\left\langle {n}_{\eta \downarrow}\left({\bf{r}}\tau\right)\right\rangle$.
Next, the contributions coming from the decouplings of the interlayer Coulomb interaction terms in Eq.(\ref{Equation_6}) have been obtained in the forms 
\begin{eqnarray}
	\delta{{\cal{S}}}_{W}=-\sum_{{\bf{r}}\sigma}\sum_{\lambda}\int^{\beta}_{0}d\tau{\Delta}^{(\lambda)}_{\sigma}{{\lambda}}^{\dag}_{\sigma}\left({\bf{r}}\tau\right){\tilde{\lambda}}_{\sigma}\left({\bf{r}}\tau\right).
	\label{Equation_18}
\end{eqnarray}
and
\begin{eqnarray}
	\delta{{\cal{H}}}_{W}=-\sum_{{\bf{r}}\sigma}\sum_{\lambda}{\Delta}^{(\lambda)}_{\sigma}{{\lambda}}^{\dag}_{\sigma}\left({\bf{r}}\tau\right){\tilde{\lambda}}_{\sigma}\left({\bf{r}}\tau\right),
	\label{Equation_19}
\end{eqnarray} 
where the summation index $\lambda$ takes two values $a$ and $b$ and ${\Delta}^{(\lambda)}_{\sigma}$ is the interlayer excitonic order parameter corresponding to the formation of the excitons between the electrons and holes at the sublattice site positions $A$-$\tilde{A}$ or $B$-$\tilde{B}$ which define the summation parameter $\lambda$. We have
\begin{eqnarray}
	{{\Delta}}^{(\lambda)}_{\sigma}=W\left\langle{{\lambda}}^{\dag}_{\sigma}\left({\bf{r}}\tau\right){\tilde{\lambda}}_{\sigma}\left({\bf{r}}\tau\right)\right\rangle.
	\label{Equation_20}
\end{eqnarray}
Then the total Hamiltonian in Eq.(\ref{Equation_1}) could be rewritten in the form which is linear in fermionic densities. In Grassmann variable notations, we have
\begin{eqnarray}
	&&{\cal{H}}_{{\rm AA}}={\cal{H}}_{\gamma_0}+{\cal{H}}_{\gamma_1}+\delta{{\cal{H}}}_{U}+\delta{{\cal{H}}}_{W}+{\cal{H}}_{V}+{{\cal{H}}}_{\rm int}+{\cal{H}}_{\mu}.
	\label{Equation_21}
	\nonumber\\
\end{eqnarray} 
In the next section, we will use this form of the Hamiltonian, for writing the expression of the inverse Green function matrix of the considered system and to derive the set of self-consistent equations.  
%
\section{\label{sec:Section_3} The particle excitation quasienergies}
%
\subsection{\label{sec:Section_3_1} The inverse Green's function and electronic band structure}
%
In the following sections, we will use total fermionic action, written in Eq.(\ref{Equation_13}), to calculate the Green's functions matrices and to derive the signle quasiparticle excitation quasienergies which form the electronic band structure in the AA-BLG system. We will pass into the reciprocal space representation for the creation and annihilation operators $\bar{\eta}_{\sigma}({\bf{r}}\tau)$ and ${\eta}_{\sigma}({\bf{r}}\tau)$. For this, we perform the Fourier transformation into the reciprocal space representation $\left({\bf{k}},\nu_{n}\right)$
\begin{eqnarray}
	{\eta}_{\sigma}({\bf{r}}\tau)=\frac{1}{\beta{N}}\sum_{{\bf{k}}\nu_{n}}\eta_{\sigma}\left({\bf{k}}\nu_{n}\right)e^{i\left({\bf{k}}{\bf{r}}-\nu_{n}\tau\right)},
	\label{Equation_22}
\end{eqnarray}
where $N$ is the number of the reciprocal lattice points and the summation in the righ hand side in Eq.(\ref{Equation_22}) is over the reciprocal wave vectors ${\bf{k}}$ and fermionic Matsubara frequencies $\nu_{n}$ with $\nu_{n}=\frac{\pi}{\beta}\left(2n+1\right)$ \cite{cite_66}, where $n$ in an integer number (i.e., $n=0,\pm1,\pm2,...$). Next, we introduce the Nambu spinors \cite{cite_63}  $\bar{\Psi}_{\sigma}\left({\bf{k}}\nu_{n}\right)$ and $\Psi_{\sigma}\left({\bf{k}}\nu_{n}\right)$ for the considered problem. We have
\begin{eqnarray} 
	{\Psi}_{\sigma}\left({\bf{k}}\nu_{n}\right)=\left(
	\begin{array}{crrrr}
		{a}_{\sigma}\left({\bf{k}}\nu_{n}\right)\\\\
		{b}_{\sigma}\left({\bf{k}}\nu_{n}\right) \\\\
		{\tilde{a}}_{\sigma}\left({\bf{k}}\nu_{n}\right) \\\\
		{\tilde{b}}_{\sigma}\left({\bf{k}}\nu_{n}\right) \\\\
	\end{array}
	\right)
	\label{Equation_23}
\end{eqnarray}
and 
\begin{eqnarray} 
	{\bar{\Psi}}_{\sigma}\left({\bf{k}}\nu_{n}\right)=\left(	\bar{{a}}_{\sigma}\left({\bf{k}}\nu_{n}\right),\bar{{b}}_{\sigma}\left({\bf{k}}\nu_{n}\right),\bar{\tilde{a}}_{\sigma}\left({\bf{k}}\nu_{n}\right), \bar{\tilde{b}}_{\sigma}\left({\bf{k}}\nu_{n}\right)\right).
	\nonumber\\
	\label{Equation_24}
\end{eqnarray}
Then, the total fermionic action could be written in the Fourier transformed form 
\begin{eqnarray}
{\cal{S}}\left[\bar{\Psi},\Psi\right]=\frac{1}{\beta{N}}\sum_{{\bf{k}}\nu_{n}}\sum_{\sigma}\bar{\Psi}_{\sigma}\left({\bf{k}}\nu_{n}\right){\cal{G}}^{-1}_{\sigma}\left({\bf{k}}\nu_{n}\right)\Psi_{\sigma}\left({\bf{k}}\nu_{n}\right).
	\nonumber\\
	\label{Equation_25}
\end{eqnarray}
Indeed, the forms of the inverse Greens functions, corresponding to the opposite spin directions, are different as we will see here, later on. In other words the total action of the system becomes composed of two parts ${\cal{S}}\left[\bar{\Psi},\Psi\right]={\cal{S}}_{\uparrow}\left[\bar{\Psi},\Psi\right]+{\cal{S}}_{\downarrow}\left[\bar{\Psi},\Psi\right]$.
For the matrices ${\cal{G}}^{-1}_{\uparrow}\left({\bf{k}}\nu_{n}\right)$ and ${\cal{G}}^{-1}_{\downarrow}\left({\bf{k}}\nu_{n}\right)$, we get the following analytical expressions
\begin{widetext}
\begin{eqnarray}
	{\cal{G}}^{-1}_{\sigma}\left({\bf{k}}\nu_{n}\right)=\left(
	\begin{array}{ccccrrrr}
		E_{1\sigma}(\nu_{n}) & -\tilde{\gamma}_{0{\bf{k}}} & -\left(\gamma_1+\Delta_{\sigma}\right) & 0\\
		-{\tilde{\gamma}_{0{\bf{k}}}}^{\ast} &E_{1\sigma}(\nu_{n})  &  0 & -\left(\gamma_1+\Delta_{\sigma}\right) \\
		-\left(\gamma_1+\Delta_{\sigma}\right) & 0 & E_{2\sigma}(\nu_{n}) & -\tilde{\gamma}_{0{\bf{k}}} \\
		0 & -\left(\gamma_1+\Delta_{\sigma}\right) & -{\tilde{\gamma}_{0{\bf{k}}}}^{\ast} & E_{2\sigma}(\nu_{n}) 
	\end{array}
	\right).
	\label{Equation_26}
\end{eqnarray}
\end{widetext}
The diagonal elements $E_{i\sigma}(\nu_{n})$ (with $i=1,2$) in the matrices Eq.(\ref{Equation_26}) represent the shifted single-particle quasienergies and are given by the following expressions
\begin{eqnarray}
	E_{i\sigma}\left(\nu_{n}\right)=-i\nu_{n}-\left(-1\right)^{i}\frac{V}{2}+x_{i\sigma}.
	\label{Equation_27}
\end{eqnarray}
In turn, the interaction-normalized parameters $x_{i\sigma}$, in Eq.(\ref{Equation_27}), are given via the expressions
\begin{eqnarray}
	&&x_{1\sigma}=2W-\mu+\left(-1\right)^{\sigma}g\gamma_0\tilde{B}+\frac{U}{2}\bar{n}_{a},
	\nonumber\\
	&&x_{2\sigma}=-\mu+\left(-1\right)^{\sigma}g\gamma_0\tilde{B}+\frac{U}{2}\bar{n}_{\tilde{a}},
	\label{Equation_28}
\end{eqnarray}
where 
\begin{eqnarray}
\left(-1\right)^{\sigma}=	\left\{
\begin{array}{cc}
	\displaystyle  & +1, \ {\rm if} \  \sigma=\uparrow,
	\newline\\
	\newline\\
	\displaystyle  & -1, \ {\rm if} \  \sigma=\downarrow.
\end{array}\right.
	\label{Equation_29}
\end{eqnarray}
Here, we have putted $\tilde{B}\equiv\mu_{\rm B}B/\gamma_0$ and the averages $\bar{n}_{a}$ and $\bar{n}_{\tilde{a}}$ signify the spin-summed average fermionic densities, at the lattice sites in the sublattices $A$ and $\tilde{A}$ (see in Fig.~\ref{fig:Fig_1}).   
Next, the parameters ${\tilde{\gamma}_{0{\bf{k}}}}$ are renormalized hopping amplitudes
\begin{eqnarray}
{\tilde{\gamma}_{0{\bf{k}}}}=\gamma_0\sum_{\bm{\mathit{\delta}}}e^{-i{{\bf{k}}\bm{\mathit{\delta}}}}.
\label{Equation_30}
\end{eqnarray}
The vectors $\bm{\mathit{\delta}}$, in Eq.(\ref{Equation_30}), represent the nearest neighbor vectors in different layers $\ell=1,2$. The components of $\bm{\mathit{\delta}}$ are the same for $\ell=1,2$ ($\bm{\mathit{\delta}}^{\ell=1}_{i}=\bm{\mathit{\delta}}^{\ell=2}_{i}\equiv \bm{\mathit{\delta}}_{i}$ with $i=1,..3$) and are given by 
\begin{eqnarray}
	\footnotesize
	\bm{\mathit{\delta}}=
	\left\{
	\begin{array}{cc}
		\displaystyle  & \bm{\mathit{\delta}}_{1}=\left(\frac{{a}}{2\sqrt{3}},\frac{a}{2}\right),
		\newline\\
		\newline\\
		\displaystyle  & \bm{\mathit{\delta}}_{2}=\left(\frac{{a}}{2\sqrt{3}},-\frac{a}{2}\right),
		\newline\\
		\newline\\
		\displaystyle  &
		\bm{\mathit{\delta}}_{3}=\left(-\frac{a}{\sqrt{3}},0\right),
	\end{array}\right.
	\label{Equation_31}
\end{eqnarray}
where $a=\sqrt{3}a_{0}$ in Eq.(\ref{Equation_31}) is the sublattice constant, while $a_{0}$ is the carbon-carbon length in the graphene layers (with $a_{0}=1.42{\AA}$). Furthermore, the expressions of spin-dependent coefficients $x_{i\sigma}$, in Eq.(\ref{Equation_28}), could be written with the help of the inverse filling coefficient $\kappa$ and average charge density difference $\delta{\bar{n}}$ between the layers. For the sublattices $A$ and $\tilde{A}$, in the AA-BLG (see in Fig.~\ref{fig:Fig_1}), these parameters could be defined in the following form
\begin{eqnarray}
	\bar{n}_{a}+\bar{n}_{\tilde{a}}=\frac{1}{\kappa},
	\nonumber\\
	\bar{n}_{\tilde{a}}-\bar{n}_{a}=\frac{\delta \bar{n}}{2}.
	\label{Equation_32}
\end{eqnarray} 
Here, the coefficient $\kappa$ describes the inverse of the total number of particles at the given lattice site position. It can be expressed with help of the filling coefficient $n_{\rm fill}$, i.e., $\kappa=1/n_{\rm fill}$. The maximum number of the parameter $\kappa$ is equal to $0.25$, which corresponds to the fully filled lattice sites. If $\kappa=0.25$ then we have the maximum average total number of particles in both layers (at the same given lattice site ${\bf{r}}$ in both layers) equal to $4$, according to the Pauli principle. If $\kappa=0.5$, then we have the half-filling case with $n_{\rm fill}=2$. In the latest case, we have only one particle per sublattice site corresponding to the fermions of type $a$ and $\tilde{a}$. The number $\delta \bar{n}$ signifies the average charge density imbalance between the layers and is defined as
\begin{eqnarray}
	\delta{\bar{n}}=\bar{n}_{2}-\bar{n}_{1},
	\label{Equation_33}
\end{eqnarray}
where the average charge densities $\bar{n}_{l}$ (with the layers indices $\ell=1,2$) mean the total average fermionic charge densities in the separate layers in the AA-BLG. Similar expressions could be written also for the sublattice charge densities $\bar{n}_{b}$ and 
$\bar{n}_{\tilde{b}}$ attached to the lattice sites $B$ and $\tilde{B}$, respectively. The average charge densities $\bar{n}_{a}$ and $\bar{n}_{\tilde{a}}$ could be expressed as
\begin{eqnarray}
	\bar{n}_{a}=\frac{1}{2}\left(\frac{1}{\kappa}-\frac{\delta{\bar{n}}}{2}\right),
	\nonumber\\
	\bar{n}_{\tilde{a}}=\frac{1}{2}\left(\frac{1}{\kappa}+\frac{\delta{\bar{n}}}{2}\right).
	\label{Equation_34}
\end{eqnarray}
Then we present the calculations results for the total single-particle quasienergies in the AA-BLG, corresponding to different spin-directions. Those quasienergies define the electronic band structure in the AA-BLG, in various interactions regimes and are the subject for further consideration in the present paper. As it hase been mentioned earlier, we consider two different spin configurations in the system, i.e., $\sigma=\uparrow$, $\sigma=\downarrow$, and corresponding secular determinants $\det{{\cal{G}}^{-1}_{\sigma}\left({\bf{k}}\nu_{n}\right)}$. The eigenvalues of the inverse Green's function matrices (for two different spin directions) could be obtained by solving the equations for secular determinants  $\det{{\cal{G}}^{-1}_{\sigma}\left({\bf{k}}\nu_{n}\right)}=0$. The solutions of those equations give the exact band structure in the AA-BLG with excitonic correlations and interaction effects considered here. For the given spin $\sigma$ we have:
\begin{widetext}
\begin{eqnarray}
	&&\epsilon_{i\sigma}\left({\bf{k}}\right)=\frac{1}{2}\left[-x_{1\sigma}-x_{2\sigma}-\left(-1\right)^{i+1}\sqrt{\left(x_{1\sigma}-x_{2\sigma}-V\right)^{2}+4\left({\tilde{\Delta}}^{2}_{\sigma}+|\tilde{\gamma}_{0{\bf{k}}}|^{2}\right)-4|\tilde{\gamma}_{0{\bf{k}}}|\sqrt{\left(x_{1\sigma}-x_{2\sigma}-V\right)^{2}+4{\tilde{\Delta}}^{2}_{\sigma}}}\right],
	\nonumber\\
	\label{Equation_35}  
\end{eqnarray}
and 
\begin{eqnarray}
	&&\epsilon_{j\sigma}\left({\bf{k}}\right)=\frac{1}{2}\left[-x_{1\sigma}-x_{2\sigma}-\left(-1\right)^{j+1}\sqrt{\left(x_{1\sigma}-x_{2\sigma}-V\right)^{2}+4\left({\tilde{\Delta}}^{2}_{\sigma}+|\tilde{\gamma}_{0{\bf{k}}}|^{2}\right)+4|\tilde{\gamma}_{0{\bf{k}}}|\sqrt{\left(x_{1\sigma}-x_{2\sigma}-V\right)^{2}+4{\tilde{\Delta}}^{2}_{\sigma}}}\right],
	\nonumber\\
	\label{Equation_36}  
\end{eqnarray}
\end{widetext}
where $i=1,2$ and $j=3,4$ and ${\tilde{\Delta}}_{\sigma}={\Delta}_{\sigma}+\gamma_1$.
Furthermore, we will show that the electronic band structure energies, given in Eqs.(\ref{Equation_35}) and $(\ref{Equation_36})$, are different for different spin orientations, due to the presence of the external magnetic field $B$. 
%
\subsection{\label{sec:Section_3_2} Self-consistent equations and total energy}
%
Indeed, for calculating numerically the electronic band structure and total energies, related to the concrete spin direction, we need to solve a set of self-consistent (SC) equations for the chemical potential ($\mu$ in our theory is a physical quantity that we calculate exactly after solving the SC equations), the average charge density difference function between the layers ($\delta\bar{n}=\bar{n}_{2}-\bar{n}_{1}$) and the excitonic order parameters $\Delta_{\uparrow}$ and $\Delta_{\downarrow}$ (those parameters are not equal $\Delta_{\uparrow}\neq\Delta_{\downarrow}$). In Appendix \ref{sec:Section_6}, we give the detailed derivation of the SC equations in the AA-BLG. Here, we just present the final form of those equations. We get
\begin{eqnarray}
	&&\frac{1}{\kappa}=\frac{1}{{N}}\sum_{{\bf{k}}}\sum^{4}_{i=1}\sum_{\sigma}\left(\alpha_{i{\bf{k}}\sigma}+\beta_{i{\bf{k}}\sigma}\right)n_{F}\left(\mu-\epsilon_{i\sigma}\left({\bf{k}}\right)\right),
	\nonumber\\
	&&\frac{\delta{\bar{n}}}{2}=\frac{1}{{N}}\sum_{{\bf{k}}}\sum^{4}_{i=1}\sum_{\sigma}\left(\beta_{i{\bf{k}}\sigma}-\alpha_{i{\bf{k}}\sigma}\right)n_{F}\left(\mu-\epsilon_{i\sigma}\left({\bf{k}}\right)\right),
	\nonumber\\
	&&\Delta_{\sigma}=\frac{W\left(\Delta_{\sigma}+\gamma_1\right)}{{N}}\sum_{{\bf{k}}}\sum^{4}_{i=1}\gamma_{i{\bf{k}}\sigma}n_{F}\left(\mu-\epsilon_{i\sigma}\left({\bf{k}}\right)\right).
	\label{Equation_37}
\end{eqnarray}
The normalization factor $N$ in Eq.(\ref{Equation_37}) is the number of the reciprocal lattice vectors ${\bf{k}}=\left(k_x,k_y\right)$. For each crystallographic direction $k_{i}$ (with $i=x,y$), we have chosen $100$ $k_{i}$-points, thus totally having a number of $N=10^{4}$ $k_{i}$-points, considered in the numerical evaluations.
The function $n_{F}\left(x\right)$ entering in the right-hand sides of the equations in Eq.(\ref{Equation_37}) is the Fermi-Dirac distribution function
\begin{eqnarray}
n_{F}\left(x\right)=1/\left(e^{\beta\left(x-\mu\right)}+1\right),
	\label{Equation_38}
\end{eqnarray}
where $\mu$ is the chemical potential in the system and should be calculated exactly, after solving the above SC equations. Particularly, the first equation in Eq.(\ref{Equation_37}) defines the chemical potential in the system (see, also, the discussion in Appendix \ref{sec:Section_6}). 
%
\begin{figure}[h!]
	\begin{center}
		\includegraphics[scale=0.53]{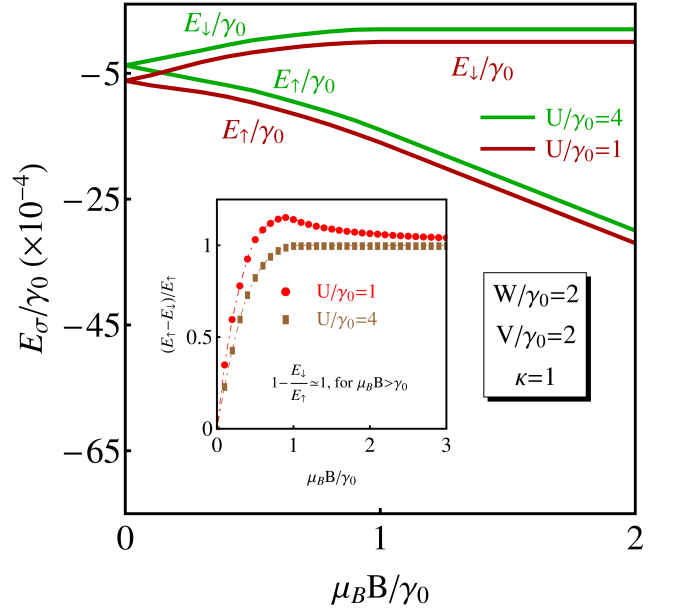}
		\caption{\label{fig:Fig_2}(Color online) Total ${\bf{k}}$-summed energies for different spin orientations $E_{\uparrow}$ and $E_{\downarrow}$ in the AA stacked bilayer graphene. We see in the picture the dependence on the magnetic field $B$. Zeeman-like field splitting was observed for the non-zero values of the external magnetic field. Different limits of the intralayer Coulomb interaction parameter have been considered, during the calculations. The lines in blue correspond to the case $U=4\gamma_0$, and the red lines correspond to the small value of the interaction parameter $U=\gamma_0$. The inverse filling coefficient $\kappa$ was set at $\kappa=1$, i.e, that is the case partial-filling. The temperature is set at $T=0$.}
	\end{center}
\end{figure} 
%
The second of equations in Eq.(\ref{Equation_37}) is the equation for density difference function $\delta{\bar{n}}$, which is defined in Eq.(\ref{Equation_37}), above. The ${\bf{k}}$-dependent parameters $\alpha_{i{\bf{k}}\sigma}$, $\beta_{i{\bf{k}}\sigma}$ and $\gamma_{i{\bf{k}}\sigma}$ are defined in the Appendix \ref{sec:Section_6}, at the end of the paper. Third equation in Eq.(\ref{Equation_37}) is the subject of two separate self-consistent equations for the excitonic order parameters $\Delta_{\uparrow}$ and $\Delta_{\downarrow}$. The system of SC equations in Eq.(\ref{Equation_37}) could be solved with the high precision by employing the finite-difference approximation method which retains the fast convergent Newton's algorithm \cite{cite_67}.

The total ${\bf{k}}$-integrated sum of the single-excitation quasienergies, in the region close to the First Brillouin Zone (FBZ), is given by the expression 
\begin{eqnarray}
	E_{\sigma}=\frac{1}{N}\sum_{{\bf{k}}}\sum^{4}_{i=1}\epsilon_{i\sigma}\left({\bf{k}}\right).
	\label{Equation_39}
\end{eqnarray}
The results are plotted in Fig.~\ref{fig:Fig_2} for both spin directions $\sigma=\uparrow$ and $\sigma=\downarrow$. We have shown, in Fig.~\ref{fig:Fig_2}, the dependence of total energies $E_{\sigma}$ on the magnetic field for different values of the normalized intralayer Coulomb interaction parameter $U/\gamma_0$. The plots in red show the total energies, for two spin directions $\sigma=\uparrow, \downarrow$ and for the small value of the Coulomb interaction parameter: $U=\gamma_0=3$ eV. The plots in blue show the total energies for large $U$ limit with $U=4\gamma_0=12$ eV. We considered, in Fig.~\ref{fig:Fig_2}, the general case of the partial-filling and we putted $\kappa=1$. The interlayer Coulomb interaction parameter is set at the value $W=2\gamma_0=6$ eV and the external electric field potential is fixed at $V=2\gamma_0=6$ eV. We observe, in Fig.~\ref{fig:Fig_2}, that at the zero value of the external magnetic field ($\tilde{B}=0$) the total energies are single-valued for both values of the Coulomb interaction parameter $U$, i.e., $E_{\uparrow}=E_{\downarrow}$. When augmenting the magnetic field $\tilde{B}\neq0$, we get the effect similar to the usual spin-Zeeman one \cite{cite_29}, and the energy bands, corresponding to different spin orientations, get split. The similar effects have been observed recently in \cite{cite_30, cite_31}, concerning the study of the magnetic field properties in the AA stacked bilayer graphene quantum dots and the energy spectrum of a magnetic quantum dot in graphene. The values of energy splitting depend on the strength of the external transverse magnetic field $B$. It is worth to mention here that each point in Fig.~\ref{fig:Fig_2} was calculated after solving the system of SC equations in Eq.(\ref{Equation_37}) (see in the next section) and by putting the obtained values of the physical parameters $\mu$, $\delta{\bar{n}}$, $\Delta_{\uparrow}$ and $\Delta_{\downarrow}$ in Eq.(\ref{Equation_38}). For $U=3$ eV, the absolute value of the total splitting between $E_{\uparrow}$ and $E_{\downarrow}$ at $\tilde{B}=1$ is equal to $\Delta{E}=|E_{\uparrow}-E_{\downarrow}|=4.8$ meV. At the higher magnetic field $\tilde{B}=3$, we get a large value for the splitting energy $\Delta{E}=14.4$ meV. We see in Fig.~\ref{fig:Fig_2} that the variation of the parameter $U$ doesn't change considerably the total energy, corresponding to given spin direction. In inset, in Fig.~\ref{fig:Fig_2}, we have calculated the ratio $\delta=\left(E_{\uparrow}-E_{\downarrow}\right)/E_{\uparrow}$. We observe that the ratio $\delta$ increases continuously with $\tilde{B}$ in the low magnetic field limit, i.e., when $\tilde{B}\in(0,1)$. At the high magnetic field values, i.e., when $\tilde{B}\geq1$, the ratio $\delta$ is stabilizing nearly to $1$, i.e., $\delta\sim 1$ and the principal contribution to the total energy $\Delta{E}$ is due to the spin direction $\sigma=\downarrow$.
%
\section{\label{sec:Section_4} Numerical Results}
%
\subsection{\label{sec:Section_4_1} Magnetic field effect}
\begin{figure}[h!]
		\begin{center}
			\includegraphics[width=7cm,height=18cm]{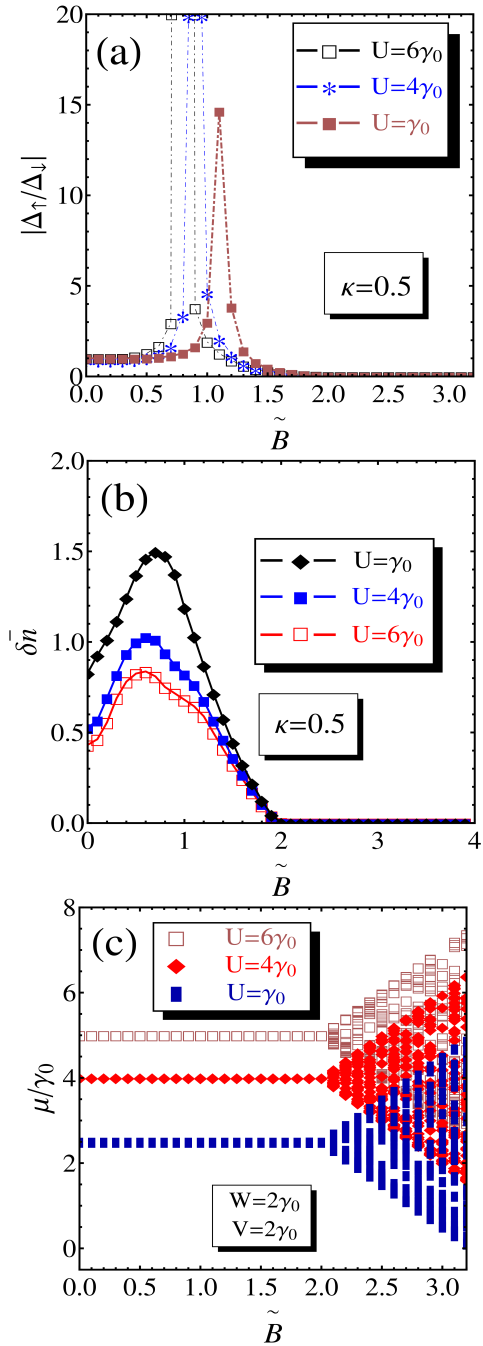}
			\caption{\label{fig:Fig_3}(Color online) The magnetic field dependence of the principal physical parameters in the AA-BLG system. The results have been obtained after solving the set of self-consistent equations, given in Eq.(\ref{Equation_37}). In different panels we have shown $\tilde{B}$-dependence (with the magnetic field parameter $\tilde{B}=\mu_{\rm B}B/\gamma_0$) of (a) the ratio $\delta=|\Delta_{\uparrow}/\Delta_{\downarrow}|$, (b) the average charge density difference function between the layers $\delta{\bar{n}}$, (c) the chemical potential. The interlayer Coulomb potential $W$ is fixed at the value $W=2\gamma_0$. The inverse filling coefficient is set at the value $\kappa=0.5$ which corresponds to the half-filling regime, and the results have been plotted for $T=0$.}
		\end{center}
	\end{figure}

%
In the present Section, we give the numerical results after solving the system of SC equations, given in Eq.(\ref{Equation_37}), at the end of the Section \ref{sec:Section_3}.
In Fig.~\ref{fig:Fig_3}, we solved the system of SC equations in Eq.(\ref{Equation_37}) for different values of the external magnetic field parameter $\tilde{B}=\mu_{\rm B}B/\gamma_0$. In the first left panel (a) in Fig.~\ref{fig:Fig_3} we have shown the $\tilde{B}$-dependence of the ratio of excitonic order parameters $\Delta_{\uparrow}$ and $\Delta_{\downarrow}$ defined as $\delta=|\Delta_{\uparrow}/\Delta_{\downarrow}|$. The interlayer Hubbard potential is fixed at the value $W=2\gamma_0$, the electric field potential is set at the value $V=2\gamma_0$, and the half-filling case is considered here with $\kappa=0.5$. It is clear from panel (a), that the $\Delta_{\uparrow}=\Delta_{\downarrow}$, for all limits of the intralayer Hubbard potential $U$, considered in the picture. We observe in panel (a) that there exists a critical value of the magnetic field parameter $\tilde{B}$, above which the excitonic gap parmeter $\Delta_{\uparrow}$ vanishes, while $\Delta_{\downarrow}\neq0$ and the maximut difference between them is observed at the intermediate values of the external magnetic field $\tilde{B}\in\left(1,1.5\right)$. Here, we realize that the mentioned critical values is equal to $\tilde{B}_{C}=2$. We see in the middle panel (b), in Fig.~\ref{fig:Fig_3}, that the average charge density difference between the layers $\delta_{n}$, which defined the charge imbalance in the AA-BLG gets vanish at the same value $\tilde{B}_{C}$ of the external magnetic field. We observe that for the small values of the external magnetic field the parameter $\delta_{\bar{n}}$ is increasing, thus the average electron population in the top layer with $\ell=2$ (see in Fig.~\ref{fig:Fig_1}, in the Section \ref{sec:Section_2}) increases with $\tilde{B}$, then it passes through a maximum (see the maximum of $\delta{\bar{n}}$ at $\tilde{B}=0.9$ when $U=\gamma_0$) and starts to decrease until the average charge neutrality (CN) occurs in the system: $\delta{\bar{n}}=0$, which means that the average electron populations in both layers become equal, i.e., $\bar{n}_{2}=\bar{n}_{1}$, and the AA-BLG system becomes charge neutral. Thus the critical value $\tilde{B_{C}}$ of the magnetic field could be called as the CN point. The vanishing of the parameter $\delta{\bar{n}}$ could be important in many aspects of the applicability of the AA-BLG system, as the nonperturbative structure device for the use in modern nano-microelectronics with the self-screened charge neutrality. Furthermore, in panel (c), in Fig.~\ref{fig:Fig_3}, we have calculated the chemical potential $\mu$ after SC equations in Eq.(\ref{Equation_37}). A very interesting degenerated behavior could be observed for the values of the magnetic field parameter $\tilde{B}$ in the interval $\tilde{B}\in(0,\tilde{B}_{C})$. When $\tilde{B}>\tilde{B}_{C}$ this degeneracy is suppressed and a huge band-solution appears at the high magnetic field values.      

	%
	\begin{figure}[h!]
		\begin{center}
			\includegraphics[width=6.8cm,height=18cm]{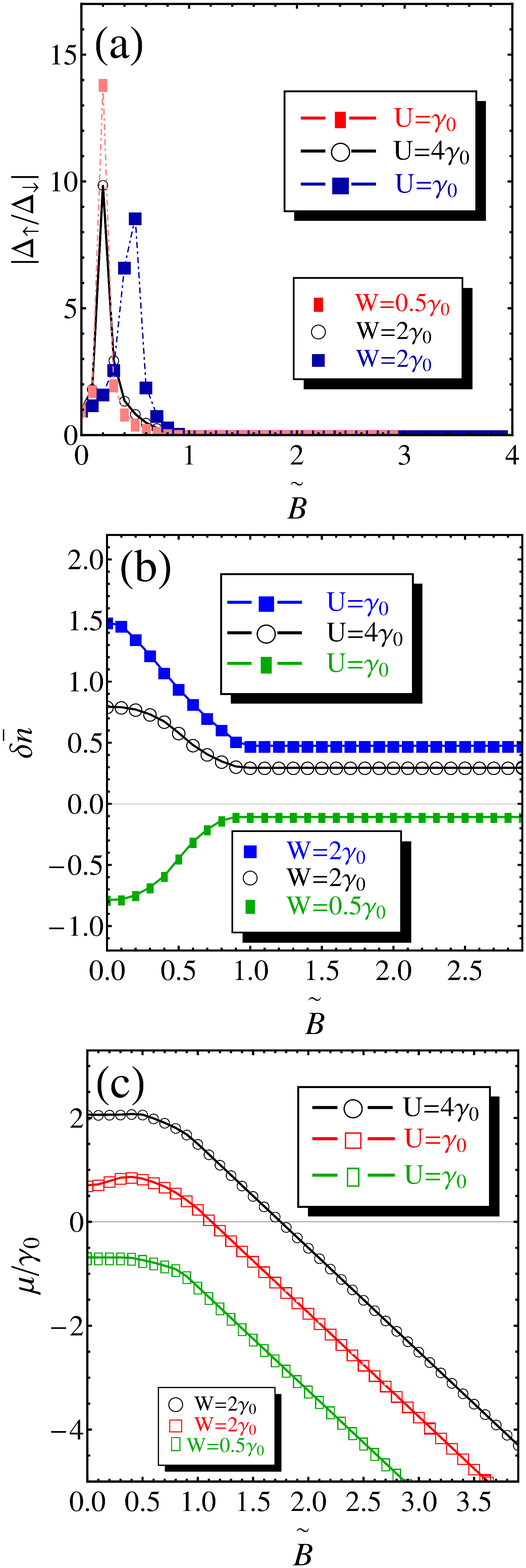}
			\caption{\label{fig:Fig_4}(Color online) The magnetic field dependence of the principal physical parameters in the AA-BLG system. The results have been obtained after solving the set of self-consistent equations, given in Eq.(\ref{Equation_37}). In different panels we have shown $\tilde{B}$-dependence (with the magnetic field parameter $\tilde{B}=\mu_{\rm B}B/\gamma_0$) of (a) the ratio $\delta=|\Delta_{\uparrow}/\Delta_{\downarrow}|$, (b) the average charge density difference function between the layers $\delta{\bar{n}}$, (c) the chemical potential. Different limits of the Hubbard interaction potentials $U$ and $W$ have been considered. The inverse filling coefficient is set at the value $\kappa=1$ which corresponds to the partial-filling regime, and the results have been plotted for $T=0$.}
		\end{center}
	\end{figure} 
%
In Fig.~\ref{fig:Fig_4}, we presented the numerical results for the same parameters, as in Fig.~\ref{fig:Fig_3}, but for the case of partial filling with $\kappa=1$. Moreover, we considered different limits of the interlayer Hubbard potential $W$. We observe in panels (a), (b) and (c) that the critical value of the magnetic field is smaller in this case with $\tilde{B}_{C}=1$. Meanwhile, from the behavior of $\delta{\bar{n}}$ in panel (b), in Fig.~\ref{fig:Fig_4}, it is clear that the CN never occurs in the limit of partial filling. Indeed, the parameter $\delta{\bar{n}}$ never vanishes in this case (see in panel (b), in Fig.~\ref{fig:Fig_4}), but takes the constant values $\delta{\bar{n}}_{0}$, when passing above the critical point $\tilde{B}_{C}$. Those constant values are different in different limits of the intralayer and interlayer Hubbard potentials and we observe a change in sign of $\delta{\bar{n}}$ and $\delta{\bar{n}}_{0}$ when passing from the high-$W$ limit to the small-$W$ limit (see, for example, the pair of plots in black and darker green). In the high-($U$,$W$) limit we have $\delta{\bar{n}}>0$ and $\delta{\bar{n}}_{0}>0$, while in the low-($U$,$W$) limit we get the negative signs $\delta{\bar{n}}<0$ and $\delta{\bar{n}}_{0}<0$. The physical meaning of observed behavior is simple: the fact that the average electron population in the top layer is higher (i.e., ${\bar{n}}_{2}>{\bar{n}}_{1}$) or smaller (${\bar{n}}_{2}<{\bar{n}}_{1}$) than the average electron population in the bottom layer depends on the strengths of the Hubbard interaction parameters $U$ and $W$. Thus, by varying the interaction potentials we can achieve the situation when the average electron population inversion takes place in the AA-BLG. Different constant values of the parameter $\delta{\bar{n}}$ at $\tilde{B}>\tilde{B}_{C}$ could also have their technological implementations as they imply that the AA-BLG system becomes a two-layer device with the desired average electron concentrations in different layers. In turn, the Hubbard interaction potentials could be tuned either by varying the applied electric field potential $V$ or by changing the interlayer separation distance $c_{0}$ in the AA-BLG. 

Another important observation in Fig.~\ref{fig:Fig_4} is related to the degenerated chemical potential $\mu$, for all values of the applied magnetic field (see in panel (c), in Fig.~\ref{fig:Fig_4}). Especially, the negative branch of the chemical potential solution at the small values of the Hubbard interaction potentials (see the plot in darker green, in panel (c), in Fig.~\ref{fig:Fig_4}) is a purposeful limit for the observation of the excitonic condensation phenomena in the AA-BLG system, i.e., when a macroscopic number of the excitonic pairs enter simultaneously into the fundamental states with the wave vector ${\bf{k}}=0$. Contrary, in the case of the half-filling regime, considered in Fig.~\ref{fig:Fig_3}, such condensation is possible for the magnetic fields up to the critical value $\tilde{B}_{C}=2$. Moreover, we see in panel (c), in  Fig.~\ref{fig:Fig_4}, that the chemical potential starts to decrease when passing through the point $\tilde{B}_{C}=1$ and it is always negative for small values of the Hubbard potentials $U$ and $W$. Due to the fact that the excitonic order parameter $\Delta_{\uparrow}=0$, for $\tilde{B}>\tilde{B}_{C}$ we conclude that excitonic condensation in the AA-BLG system (with $\mu<0$) is due principally to the spin direction $\sigma=\downarrow$ visa vie the excitonic order parameter $\Delta_{\downarrow}$, remaining in this case (see in panel (a), in Fig.~\ref{fig:Fig_4}).  
%
\subsection{\label{sec:Section_4_2} Average charge density difference and interaction potentials}
%
\begin{figure}[h!]
	\begin{center}
		\includegraphics[scale=0.255]{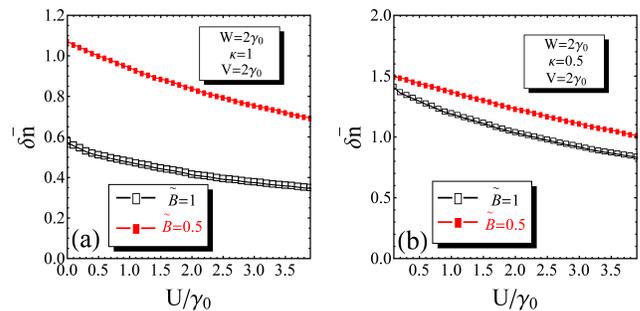}
		\caption{\label{fig:Fig_5}(Color online) The average charge density difference $\delta{\bar{n}}$, as a function of the intralayer Coulomb interaction parameter $U$. The results have been obtained after solving the set of self-consistent equations, given in Eq.(\ref{Equation_37}). In panels (a) and (b) we have shown the behavior of the function $\delta{\bar{n}}$ in the limit of partial-filling  with $\kappa=1$ (see in (a)) and half-filling $\kappa=0.5$ (see in (b)). Different values of the magnetic field parameter $\tilde{B}$ have been considered in both panels. The interlayer Coulomb interaction potential $W$ and the electric field potential $V$ are set at the values $W=V=2\gamma_0$. The results have been obtained for $T=0$.}
	\end{center}
\end{figure} 
%
\begin{figure}[h!]
	\begin{center}
		\includegraphics[scale=0.255]{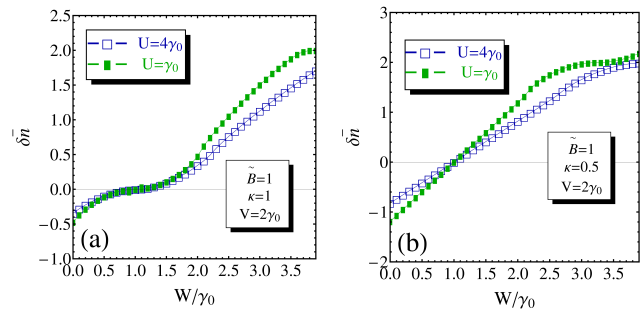}
		\caption{\label{fig:Fig_6}(Color online) The average charge density difference $\delta{\bar{n}}$, as a function of the interlayer Coulomb interaction parameter $W$. The results have been obtained after solving the set of self-consistent equations, given in Eq.(\ref{Equation_37}). In panels (a) and (b) we have shown the behavior of the function $\delta{\bar{n}}$ in the limit of partial-filling  with $\kappa=1$ (see in (a)) and half-filling $\kappa=0.5$ (see in (b)). Different values of the interaction potential $U$ have been considered in both panels. The electric field potential $V$ is set at the value $V=2\gamma_0$. The magnetic field parameter is fixed at the value $\tilde{B}=1$ and the results have been plotted for $T=0$.}
	\end{center}
\end{figure} 
%
In Figs.~\ref{fig:Fig_5} and ~\ref{fig:Fig_6}, we have shown the local Hubbard interactions effects on the average charge density difference $\delta{\bar{n}}$, for two different values of the magnetic field parameter $\tilde{B}$. Particularly, in Fig.~\ref{fig:Fig_5}, we give the numerical results for $\delta{\bar{n}}$ as a function of intralayer Hubbard potential $U$. The interlayer Hubbard potential is set at the value $W=2\gamma_0$ and the external gate potential is set at the value $V=2\gamma_0$. For both filling regimes, pres

The partial-filling and half-filling regimes have been considered in Fig.~\ref{fig:Fig_5} with $\kappa=1$ in panel (a) and $\kappa=0.5$ in panel (b). In both cases, we observe a nearly linear decrease of $\delta{\bar{n}}$ as a function of $U$ and the average charge imbalance between the layers is larger in the case of a small magnetic field (see the plots in red, corresponding to the magnetic field $\tilde{B}=0.5$). The general observation in Fig.~\ref{fig:Fig_5} is that the intralayer Hubbard-$U$ potential is stabilizing the average charge imbalance between the layers, nevertheless, the CN never occurs in this case, even for the reasonably high values of the parameter $U$. 

Furthermore, in Fig.~\ref{fig:Fig_6}, we have calculated the average charge density difference $\delta{\bar{n}}$ as a function of the interlayer Hubbard-$W$ interaction potential. Again, the partial-filling (see in panel (a)) and half-filling (see in panel (b)) regimes have been considered and the magnetic field is set at the value $\tilde{B}=1$. The external gate potential is fixed at the value $V=2\gamma_0$. We see that at the intermediate values of the interlayer Coulomb potential $W_{0}$ (which depends on the parameter $\kappa$) the function $\delta{\bar{n}}$ crosses the $W$-axis and $\delta{\bar{n}}=0$.  
It is remarkable to note that the crossing point is the same for both strong and weak Hubbard-$U$ interaction limits and the function $\delta{\bar{n}}$ changes its sign when passing through the CN point. Thus, by varying the interlayer Coulomb potential we can tune the CN limit for the function $\delta{\bar{n}}$ and also to achieve the average electron population reversion by the change of sign of the function $\delta{\bar{n}}$, i.e., $\delta{\bar{n}}<0$ for $W<W_{0}$ and $\delta{\bar{n}}>0$, for $W>W_{0}$. It is also important to notice here that the function $\delta{\bar{n}}$ is always increasing with $W$ which means that the high values of $W$ bring the system out of the CN equilibrium position.  
%
\subsection{\label{sec:Section_4_3} The excitonic condensation and magnetic field}
%

\begin{figure}[h!]
		\begin{center}
			\includegraphics[width=7cm,height=17cm]{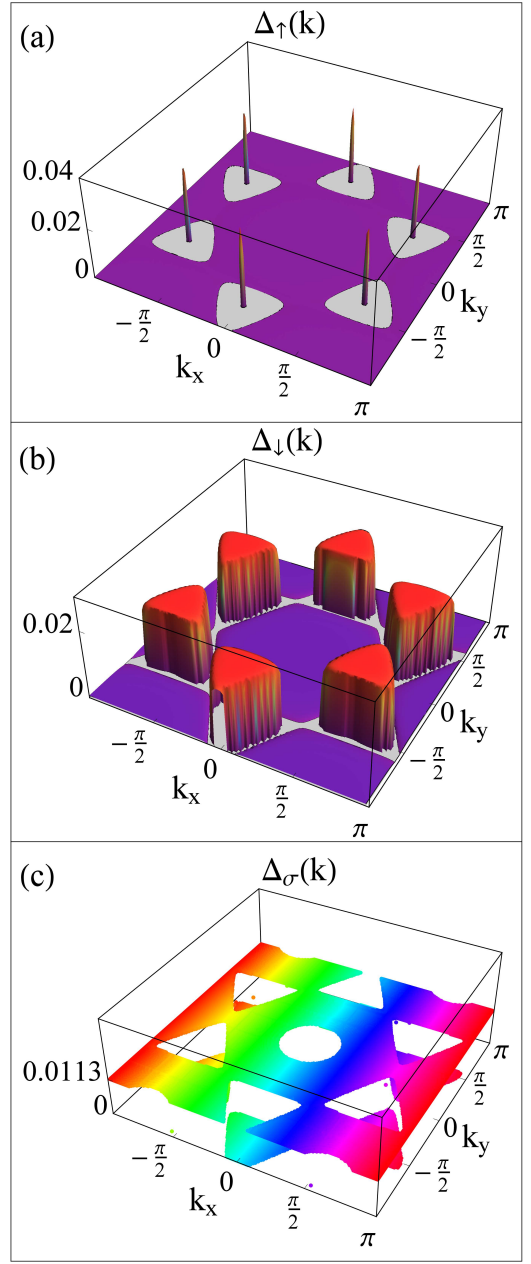}
			\caption{\label{fig:Fig_7}(Color online) The total ${\bf{k}}$-map for the excitonic order parameter $\Delta_{\sigma}\left({\bf{k}}\right)$. The plots are shown for both spin directions $\sigma=\uparrow, \downarrow$ and for the zero value of the magnetic field $\tilde{B}=0$. The other parameters are set as $U=\gamma_0$, $\kappa=0.5$ and $W=\gamma_0$ (see in (a)), $U=\gamma_0$, $\kappa=0.5$ and $W=2\gamma_0$ (see in (b)) and $U=\gamma_0$, $\kappa=1$ and $W=2\gamma_0$ (see in (c)). The large white-sockets are formed in (a) which join each other in the form of hexagonal star-like hole-pocket (see in (b)) when increasing $W$. In the middle of the hole-star vertices (see in (b)), large exciton condensate peaks appear. The peaks coincide for different spin orientations. The plot for the case away from the half-filling (this is the case of the partial-filling with $\kappa=1$) is shown in (c). The temperature is set at $T=0$.}
		\end{center}
	\end{figure}

In Fig.~\ref{fig:Fig_7} we have shown the ${\bf{k}}$-map for the excitonic order parameters $\Delta_{\uparrow}\left({\bf{k}}\right)$ and $\Delta_{\downarrow}\left({\bf{k}}\right)$ at the zero value of the external magnetic field, i.e., $\tilde{B}=0$. The half-filling case was considered during the numerical evaluations corresponding to the inverse filling parameter $\kappa=0.5$. The other parameters are set as it was shown in the figure, i.e., $W=U=\gamma_0=3$ eV and $\kappa=0.5$ (see in the panel (a)), $U=\gamma_0=3$ eV, $W=2\gamma_0$ and $\kappa=0.5$(see in the panel (b)) and $U=\gamma_0=3$ eV, $W=2\gamma_0$ and $\kappa=1$ (see in the panel (c)). 
%
%
The external electric field potential is set at the value $V=2\gamma_0=6$ eV. We see that there are condensate peaks in the middle of the hexagonal sockets, formed in the reciprocal ${\bf{k}}$-space for $W=\gamma_0$. When augmenting the interlayer interaction parameter (see in the middle panel (b), in Fig.~\ref{fig:Fig_7}) the hexagonally-arranged sockets are larger in this case and they stick together by forming a star-like (with six vertices) pocket-topology in the plane $(k_x, k_y)$. Moreover, in the middle of the star-like vertices, the large and dense exciton condensate peaks appear (see in the panel (b), in Fig.~\ref{fig:Fig_7}). We observe also that the excitonic order parameters corresponding to different spin orientations are equal, i.e., $\Delta_{\uparrow}=\Delta_{\downarrow}$. Therefore, the interlayer Coulomb interaction influences considerably and contributes to the formation of the excitonic pairs and condensate states. In the right panel (c), in Fig.~\ref{fig:Fig_7}, we have evaluated the parameters $\Delta_{\uparrow}$, and $\Delta_{\downarrow}$, away from the half-filling regime and we put $\kappa=1$. The interaction parameter $W$ is set at the value $W=2\gamma_0$ as in the case of figure (B). We see in (c) that the holly star-like pocket topology is present in that case. Moreover, the excitonic pairing states are not zero in this limit and form a surface in the reciprocal space. The change in the order parameter $\Delta_{\sigma}$ from the peak-like condensate structure to the surface of pairing could be attributed to the partial-filling considered in (c) (with $\kappa=1$). 

		
	\begin{figure}[h!]
		\begin{center}
			\includegraphics[width=7cm,height=17cm
			]{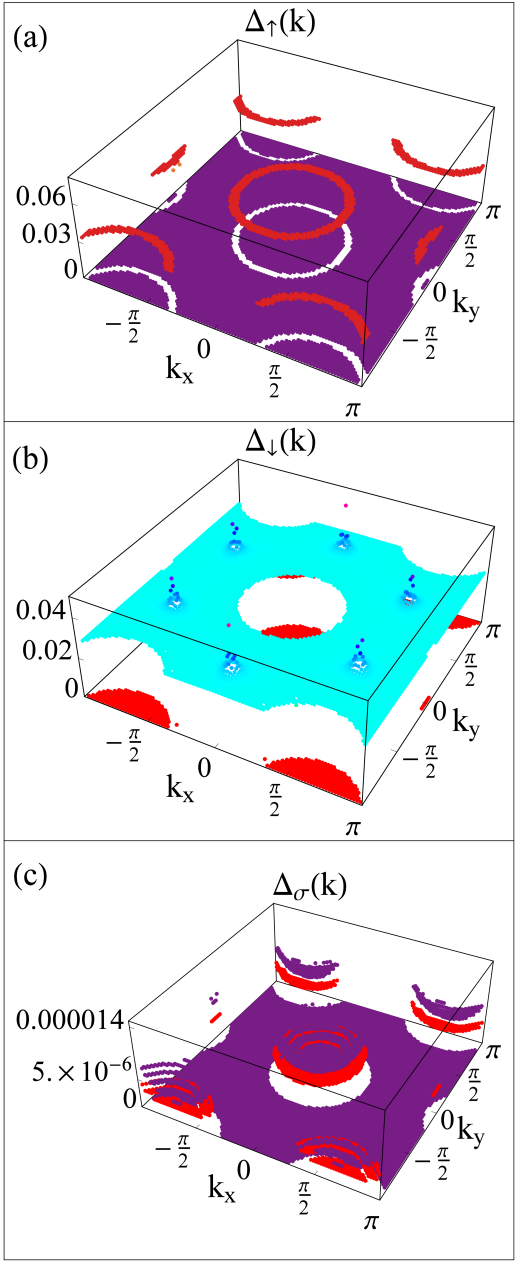}
			\caption{\label{fig:Fig_8}(Color online) The total ${\bf{k}}$-map for the excitonic order parameter $\Delta_{\sigma}\left({\bf{k}}\right)$. The plots are shown for both spin directions $\sigma=\uparrow, \downarrow$ and for the magnetic field $\tilde{B}=1$. The other parameters are set as $U=\gamma_0$, $W=\gamma_0$ and $V=2\gamma_0$. Half-filling case is considered in the calculations, i.e., $\kappa=0.5$ (one particle per-site). The figures from left to right ((a)-(c)) show the ${\bf{k}}$-map for $\Delta_{\uparrow}\left({\bf{k}}\right)$ (a), $\Delta_{\downarrow}\left({\bf{k}}\right)$ ((b)) and the mixture of them ($\Delta_{\uparrow}\left({\bf{k}}\right)$,$\Delta_{\downarrow}\left({\bf{k}}\right)$) (see in (c), the low-energy cutoff of previous images). The excitonic order parameters are not the same for different spin-orientations and exciton condensate peaks appear only in the case of $\sigma=\downarrow$ (see the peaks on the light-blue background of the excitonic pair formation surface, in  (b)). The temperature is set at $T=0$.}
		\end{center}
	\end{figure} 

\par

In Fig.~\ref{fig:Fig_8}, we give the ${\bf{k}}$-map for $\tilde{B}=1$ and for $W=U=\gamma_0=3$ eV. We see that the magnetic field induces the differences between the excitonic order parameters $\Delta_{\uparrow}$ and $\Delta_{\downarrow}$ and $\Delta_{\uparrow}\neq \Delta_{\downarrow}$, in this case (see in panels (a) and (b), in Fig.~\ref{fig:Fig_8}). Moreover, there exist the excitonic pairings with low and high energy states in this case which are shown in panels (a), (b) (with high energy excitonic states) and (c) (with low energy excitonic states). Particularly, in panel (a), the high energy excitonic states appear for $\sigma=\uparrow$ in the form of separated red rings in the ${\bf{k}}$-space. This topology of the excitonic pairing is furthermore transformed into the entire surface-like pairing topology for $\sigma=\downarrow$. Moreover the excitonic condensate peaks appear on the light-blue background surface of excitonic pairing states with $\sigma=\downarrow$ (see in panel (b), in Fig.~\ref{fig:Fig_8}). These high-energy excitonic condensate peaks are more intense than the similar peaks in the case $\tilde{B}=0$ (see in panel (b)). In panel (c), in Fig.~\ref{fig:Fig_8}, we have shown the excitonic order parameters $\Delta_{\uparrow}$ and $\Delta_{\downarrow}$, in the low-energy scale. We see that the topological disk-like regions in pink (see in (a), for $\Delta_{\uparrow}$) and red (see in (b), for $\Delta_{\downarrow}$) appear in the form of two "embedded-cups" (concavities), in the right panel (c), in Fig.~\ref{fig:Fig_8}. It follows, from the picture in (c) that $\Delta_{\uparrow}\left({\bf{k}}\right)>\Delta_{\downarrow}\left({\bf{k}}\right)$, at the low-energy scale.
	
		\begin{figure}[h!]
		\begin{center}
			\includegraphics[width=7cm,height=17cm
			]{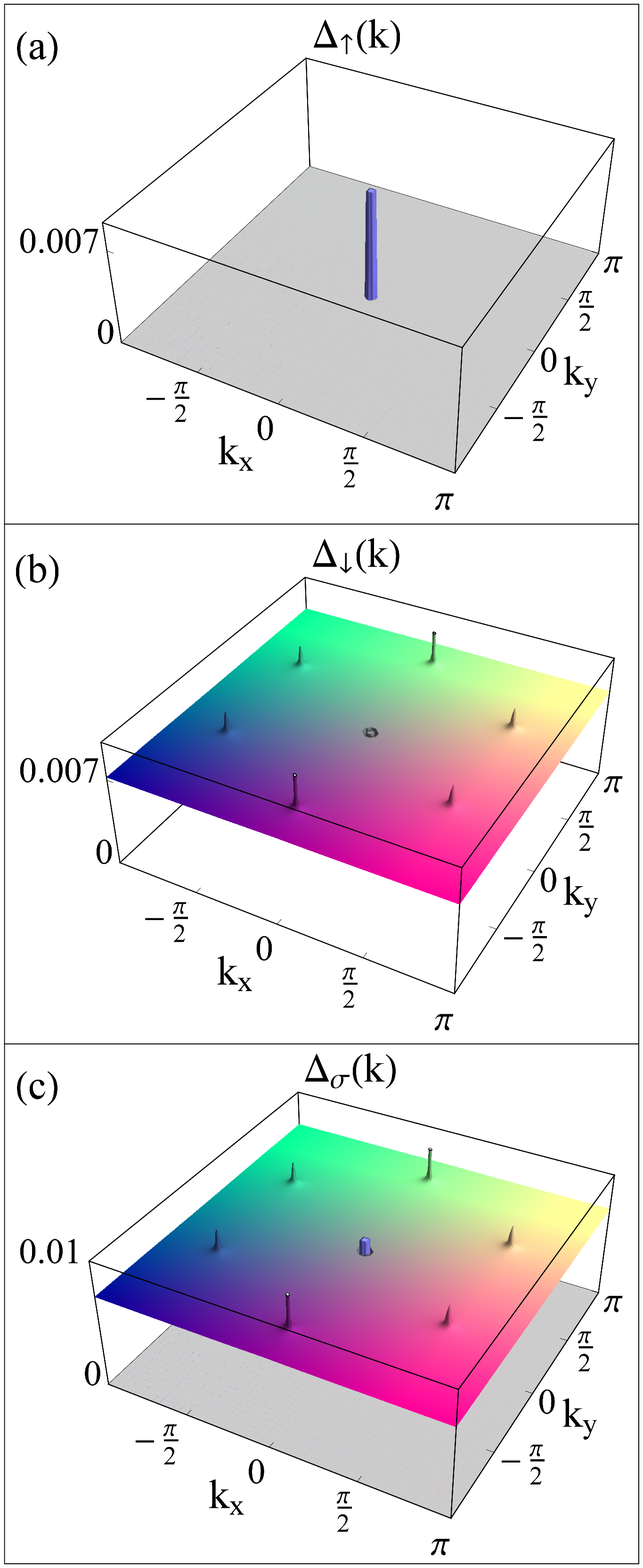}
			\caption{\label{fig:Fig_9}(Color online) The total ${\bf{k}}$-map for the excitonic order parameter $\Delta_{\sigma}\left({\bf{k}}\right)$. The plots are shown for both spin directions $\sigma=\uparrow, \downarrow$ and for the high magnetic field value $\tilde{B}=2$. The other parameters are set as $U=\gamma_0$, $W=\gamma_0$ and $V=2\gamma_0$. Half-filling case is considered in the calculations, i.e., $\kappa=0.5$ (one particle per-site). The figures from left to right (see in panels (a)-(c)) show the ${\bf{k}}$-map for $\Delta_{\uparrow}\left({\bf{k}}\right)$ (see in (a)), $\Delta_{\downarrow}\left({\bf{k}}\right)$ (see in (b)) and the mixture of them ($\Delta_{\uparrow}\left({\bf{k}}\right)$,$\Delta_{\downarrow}\left({\bf{k}}\right)$) (see in (c)). The excitonic gap parameters are not the same for different spin-orientations and an islolated exciton condensate peak appear in (a) for $\sigma=\uparrow$ at the origin of the ${\bf{k}}$-space $|{\bf{k}}|=0$. The excitonic pair formation regions are surrounded by the strongly pronounced condensate peaks for $\sigma=\downarrow$ (see in panel (b)) and a hole remains at ${\bf{k}}=0$ for $\Delta_{\downarrow}$. The temperature is set at $T=0$.}
		\end{center}
	\end{figure} 

In Fig.~\ref{fig:Fig_9} , we have shown the ${\bf{k}}$-map for the excitonic order parameters at the high magnetic field limit with $\tilde{B}=2$ at the half-filling case with $\kappa=0.5$. In the left panel (a), the ${\bf{k}}$-dependence is shown for the excitonic gap parameter $\Delta_{\uparrow}\left({\bf{k}}\right)$. The interlayer Coulomb interaction is set at the value $W=2\gamma_0$. We see, in Fig.~\ref{fig:Fig_9}, that a single, large excitonic condensate peak appears at the origin of the Brillouin Zone, i.e., at the value of the reciprocal wave vector $|{\bf{k}}|=0$. This isolated pure-condensate peak shows clearly that the excitonic condensation is possible in AA BLG at the high magnetic field regime. Furthermore, in panel (b), we see that, for the opposite spin direction ($\sigma=\downarrow$), the excitonic condensate islands appear and form a hexagonal like lattice structure in the reciprocal ${\bf{k}}$-space, surrounded by the sea of the excitonic pairs. Next, a hole pocket appears at the origin $|{\bf{k}}|=0$ (i.e., at the place of the condensate peak shown in panel (a)). In the right panel (c), in Fig.~\ref{fig:Fig_9}, we have shown both order parameters $\left(\Delta_{\uparrow}\left({\bf{k}}\right),\Delta_{\downarrow}\left({\bf{k}}\right)\right)$ and we see how the excitonic condensate peak, observed in (a), is merging with the excitonic pair formation surface (see in panel (b)), exactly at the origin $|{\bf{k}}|=0$ of the reciprocal space. 
         
%
\begin{figure}[h!]
	\begin{center}
		\includegraphics[width=8.5cm,height=19.5cm]{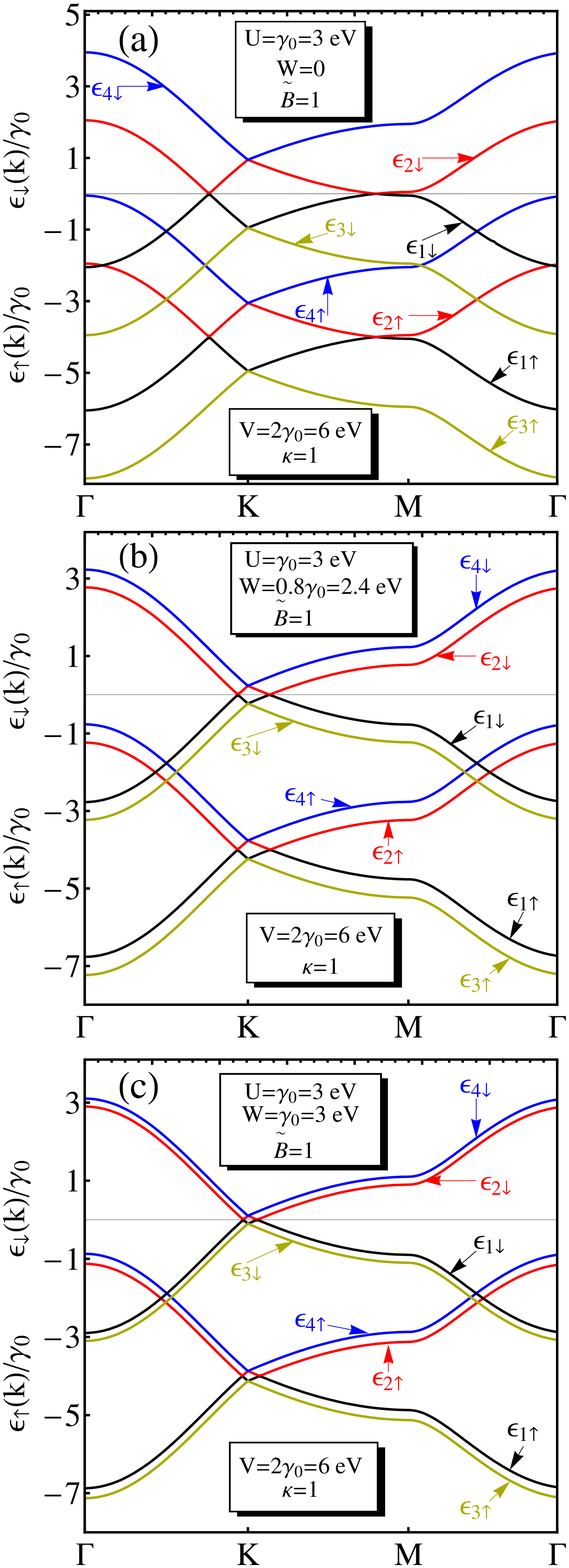}
		\caption{\label{fig:Fig_10}(Color online) The electronic band structure in the AA bilayer graphene, for different values of the interlayer Coulomb interaction parameter $W$ (see in panels (a)-(c)). The magnetic field is fixed at the value $\tilde{B}=1$ and the partial-filling is considered in the layers with $\kappa=1$. Eight different energy bands are shown in the pictures $\epsilon_{\uparrow}({\bf{k}}) \cup\epsilon_{\downarrow}({\bf{k}})$, (see colored arrows, near each energy band with the given spin direction). The temperature is set at $T=0$.}
	\end{center}
\end{figure} 
%
\subsection{\label{sec:Section_4_4} Band-structure}
%
%
\subsubsection{\label{sec:Section_4_4_1} The role of the interlayer coupling}
%

Hereafter, we present the results for the electronic band structure in the AA-BLG system. The various values of the interlayer Coulomb interaction parameter $W$ have been considered in Fig.~\ref{fig:Fig_10}. For the other physical parameters in the system we have chosen the following values $\tilde{B}=\mu_{B}B/\gamma_0=1$, $U=\gamma_0=3$ eV and $\kappa=1$ (the case of the partial-filling in the layers). The external electric gate potential is set at $V=2\gamma_0=6$ eV. The results plotted in Fig.~\ref{fig:Fig_10} have been obtained for three different values of the interaction potential $W$: $W=0$ (see in top panel (a)), $W=0.8\gamma_0=2.4$ eV (see in middle panel (b)) and $W=U=\gamma_0=3$ eV (see in bottom panel (c)). The band structure for both spin directions $\sigma=\uparrow,\downarrow$ have been shown in the picture. To distinguish different energy branches ($8$, in our case), in Fig.~\ref{fig:Fig_10}, which correspond to different spin orientations $\sigma=\uparrow, \downarrow$, we have putted small arrows near each energy band $\epsilon_{i\sigma}\left({\bf{k}}\right)$ (with $i=1,...4$). We see that for the non-interacting layers, i.e., when $W=0$ (see in panel (a), in Fig.~\ref{fig:Fig_10}), we have a large displation between the band structures corresponding to different spin directions $\sigma=\uparrow$ and $\sigma=\downarrow$. This effect of splitting is caused by the influence of the external magnetic field $B$. Furthermore, this displation is narrowing for the non-zero values of the parameter $W$ (see in panels (b) and (c)). Those band structures (for $\sigma=\uparrow$ and $\sigma=\downarrow$) get the very close when $W=U=\gamma_0=3$ eV (see in panel (c), in Fig.~\ref{fig:Fig_10}). Thus when $U=W$ (this is the most realistic case when estimating the scales of Coulomb interactions in BLG structures) the region embedded between the linearly crossing energy bands gets very small. Additionally, we obtain a doubled number of such embedded regions (see in combined view $\epsilon_{\uparrow}({\bf{k}}) \cup\epsilon_{\downarrow}({\bf{k}})$) due to the linear crossing (with no possible band-gap) of the energy bands, corresponding to different spin directions. For all considered values of the interlayer Coulomb interaction parameter $W$, we have the intersections $\epsilon_{2\uparrow}\cap\epsilon_{1\downarrow}$, $\epsilon_{2\uparrow}\cap\epsilon_{3\downarrow}$, $\epsilon_{4\uparrow}\cap\epsilon_{1\downarrow}$, $\epsilon_{4\uparrow}\cap\epsilon_{3\downarrow}$. It is worth to mention that those additional regions, embedded in the middle of the intersection points are gapless for all values of the parameter $W$ and external magnetic field $B$. 
%
%
\begin{figure}[h!]
	\begin{center}
		\includegraphics[scale=0.28]{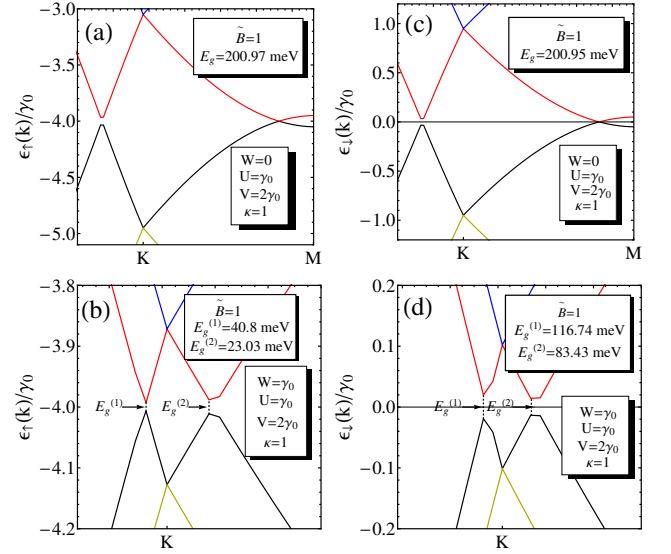}
		\caption{\label{fig:Fig_11}(Color online) The electronic band structure in the AA bilayer graphene, for different values of the interlayer Coulomb interaction parameter $W$ (see in panels (a)-(f)). The magnetic field is fixed at the value $\tilde{B}=1$ and the partial-filling is considered in the layers with $\kappa=1$. Four different energy bands are shown in different panels corresponding to each spin direction $\sigma=\uparrow$ or $\downarrow$. The results of the band structure in left panels (a), (c) and (e) correspond to $\sigma=\uparrow$ and the results in right panels (b), (d) and (f) correspond to $\sigma=\downarrow$. The narrow-range energy structure was shown in the panels, which demonstrates the band-gap opening in the electronic band structure. The band-gap opening in is due by the effect of the interlayer Coulomb interaction $W$. The temperature is set at $T=0$. }
	\end{center}
\end{figure} 
%
In panels (a)-(d) in Fig.~\ref{fig:Fig_11} the electronic band structure at the narrow regions of $\epsilon_{\uparrow}\left({\bf{k}}\right)$ and $\epsilon_{\downarrow}\left({\bf{k}}\right)$, where the band-gap is opening in the system. The values of the physical parameters in the system are the same like in Fig.~\ref{fig:Fig_10}. We see in the panels (a) and (c) that at $W=0$ a very large single band-gap $E_{g}$ is opening in the system of order $E_{g}=200.97$ meV, for $\sigma=\uparrow$ in the direction $\Gamma\rightarrow K$ on the $|{\bf{k}}|$-axis and $E_{g}=200.95$ meV for $\sigma=\downarrow$. 

The value of the band-gap for $\sigma=\uparrow$ is decreasing drastically for $W=0.8\gamma_0=2.4$ eV (with slightly displaced value of $|{\bf{k}}|$-point in the direction $\Gamma\rightarrow K$) and we have $E^{(1)}_{g}=67.38$ meV (see in panel (b), in Fig.~\ref{fig:Fig_11}). Nevertheless, the value of the band-gap still very large for the spin direction $\sigma=\downarrow$, and we have $E_{g}=273.95$ meV. Moreover, for $\sigma=\uparrow$, a second smaller gap is opening in the direction $K\rightarrow M$ with the value $E^{(2)}_{g}=45.6$ meV. Indeed, this second gap in the direction $K\rightarrow M$ is totally absent for the spin direction $\sigma=\downarrow$ (see in panel (e), in Fig.~\ref{fig:Fig_11}).  

In panels (b) and (d), in Fig.~\ref{fig:Fig_11}) we considered the large value of the interlayer interaction parameter $W$, and we calculated the band structure for $W=\gamma_0=3$ eV.
The value of the band-gap for $\sigma=\uparrow$ is decreasing drastically for $W=\gamma_0$ attaining the value of order of $E^{(1)}_{g}=40.8$ meV (see in panel (b), in Fig.~\ref{fig:Fig_11})), while in the channel $\sigma=\downarrow$ the band-gap still very large and we get $E^{(1)}_{g}=116.74$ meV (see in panel (d), in Fig.~\ref{fig:Fig_11})). Moreover, for both spin channels $\sigma=\uparrow$ and $\sigma=\downarrow$, a second smaller gap $E^{(2)}_{g}$ is opening in the direction $K\rightarrow M$ with the value $E^{(2)}_{g}=23.03$ meV for $\sigma=\uparrow$ for the spin channel $\sigma=\uparrow$ and we obtain a very large value for the second gap for $\sigma=\downarrow$ with $E^{(2)}_{g}=83.43$ meV.

The band-gap $E^{(2)}_{g}$ for $\sigma=\downarrow$ is greater from the value of $E^{(1)}_{g}$ for $\sigma=\uparrow$ nearly by a factor of two. It is particularly worth to mention that the distance $|\Delta{\bf{k}}|$ between the $|{\bf{k}}|$-points at which those two gaps are opening is narrowing for the large values of $W$ and this observation is true for both spin-directions $\sigma=\uparrow$ and $\sigma=\downarrow$. Due to the narrow energy range of excitations, those energy band-gaps could be observed experimentally by fast light emitting photon-sources such as X-ray lasers \cite{cite_68}. Another important observation is related to the induced large excitation gap $E^{(2)}_{g}$ (with the value of the gap $E^{(2)}_{g}=83.43$ meV) in the energy spectrum of $\epsilon_{\downarrow}\left({\bf{k}}\right)$, in the direction $K\rightarrow M$ (see in panel (d), in Fig.~\ref{fig:Fig_11}) and nearly flat band-energy regions at the large value of $W$. Indeed, surprisingly, a sufficiently large ${\bf{k}}$-space region is found in the band structure corresponding to the direction of spin opposite to the magnetic field ($\sigma=\downarrow$), where the usually crossing bands along direction $K\rightarrow M$ get largely separated and flattened near the zero value of the energy axis (see in panel (d), in Fig.~\ref{fig:Fig_11}). The general observation resulting from Fig.~\ref{fig:Fig_11} is that the interlayer Coulomb interaction $W$ plays a destructive role on the band-gaps $E^{(1)}_{g}$ for $\sigma=\uparrow$ (see in panels (a), (c)). Thus, we have demonstrated in Fig.~\ref{fig:Fig_11}, that the increase of the interlayer interaction potential at the fixed non-zero value of the external magnetic field (experimentally, the changes in $W$ could be realized by changing the separation distance between the layers \cite{cite_36} in the AA-BLG), has a destructive effect on the band-gaps in the system and a phase transition from semiconducting to the weak-metallic state occurs in this case for the spin channel $\sigma=\uparrow$. This result is very similar with the results in \cite{cite_36}.   
%
\subsubsection{\label{sec:Section_4_4_2} The role of the magnetic field}
\begin{figure}[h!]
	\begin{center}
		\includegraphics[scale=0.295]{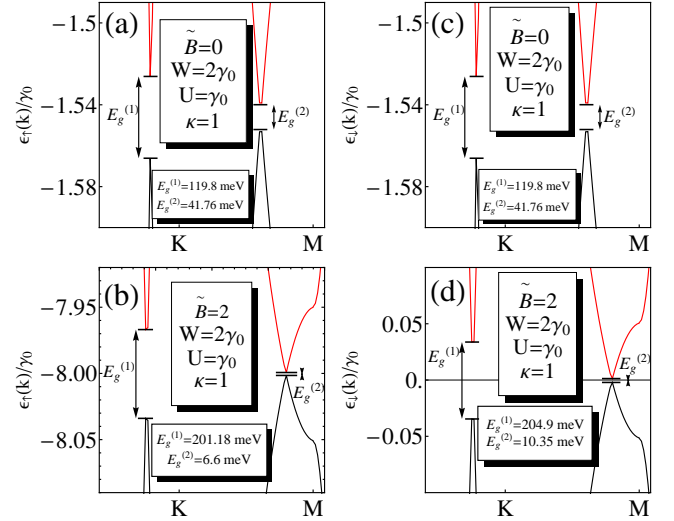}
		\caption{\label{fig:Fig_12}(Color online) The electronic band structure in the AA bilayer graphene, for different values of the external magnetic field $\tilde{B}$ (see in panels (a)-(f)). The narrow-range energy structure was shown in panels, where we presented the principal regions on the $|{\bf{k}}|$-axis at which the band-gap is opening in the AA BLG. The band-gaps in panels (a) and (d) at $\tilde{B}=0$ are due to the effect of the interlayer interaction parameter, which is fixed at the value $W=2\gamma_0=6$ eV. The partial-filling case is considered with $\kappa=1$, and the temperature is set at $T=0$.}
	\end{center}
\end{figure}

%
In Fig.~\ref{fig:Fig_12} we examine the role of the external magnetic field $\tilde{B}$ on the electronic band structure and energy gaps in the AA BLG and we considered both low (see in panels (a) and (c)) and high magnetic field limits (see in panels (b) and (d)). We fixed the interaction parameter $W$ to a sufficiently high value $W=2\gamma_0$, and the partial-filling case is considered with the inverse filling coefficient $\kappa=1$. The temperature is set at the value $T=0$ and the intralayer Coulomb interaction is set at the value $U=\gamma_0$.
%
%
%
We see that even at $\tilde{B}=0$ there exist two gaps $E^{(1)}_{g}$ and $E^{(2)}_{g}$ in both energy spectrum $\epsilon_{\uparrow}\left({\bf{k}}\right)$
and $\epsilon_{\downarrow}\left({\bf{k}}\right)$ (see in panels (a) and (c), in Fig.~\ref{fig:Fig_12}) and they are exactly the same due to the spin-symmetry in the case and the absence of the Zeeman splitting. We obtain $E^{(1)}_{g}\left(\sigma=\uparrow\right)=E^{(1)}_{g}\left(\sigma=\downarrow\right)=119.8$ meV and $E^{(2)}_{g}\left(\sigma=\uparrow\right)=E^{(2)}_{g}\left(\sigma=\downarrow\right)=41.76$ meV). 
The second small gap $E^{(2)}_{g}$ is comparable with the values of the band-gap obtained via SOC interaction effects \cite{cite_17, cite_18, cite_19, cite_20}. Next, when increasing the magnetic field parameter $\tilde{B}$, given in panels (b) and (d) in Fig.~\ref{fig:Fig_12}), the band-gaps $E^{(1)}_{g}$ for $\sigma=\uparrow,\downarrow$ get increased considerably, by approaching with their values to the band-gap energies in the usual direct band-gap semiconductors of type A$^{\rm III}$B$^{\rm IV}$ or A$^{\rm IV}$B$^{\rm IV}$ \cite{cite_48}. Namely, we have $E^{(1)}_{g}\left(\sigma=\uparrow\right)=201.18$ meV, $E^{(2)}_{g}\left(\sigma=\downarrow\right)=204.9$ meV. The energy band-gaps $E^{(2)}_{g}$ decrease drastically for both spin channels and we obtain very small values, of order of $E^{(2)}_{2}=6.6$ meV, for $\sigma=\uparrow$ and, for $\sigma=\downarrow$, we observe that $E^{(2)}_{g}=10.35$ meV. It is worth to mention that the energy band-gaps $E^{(2)}_{2}$ obtained here still much larger than the band-gaps obtained in \cite{cite_17,cite_18, cite_19, cite_20} when considering the spin-orbit coupling only.  
%
%
\begin{figure}[h!]
	\begin{center}
		\includegraphics[scale=0.26]{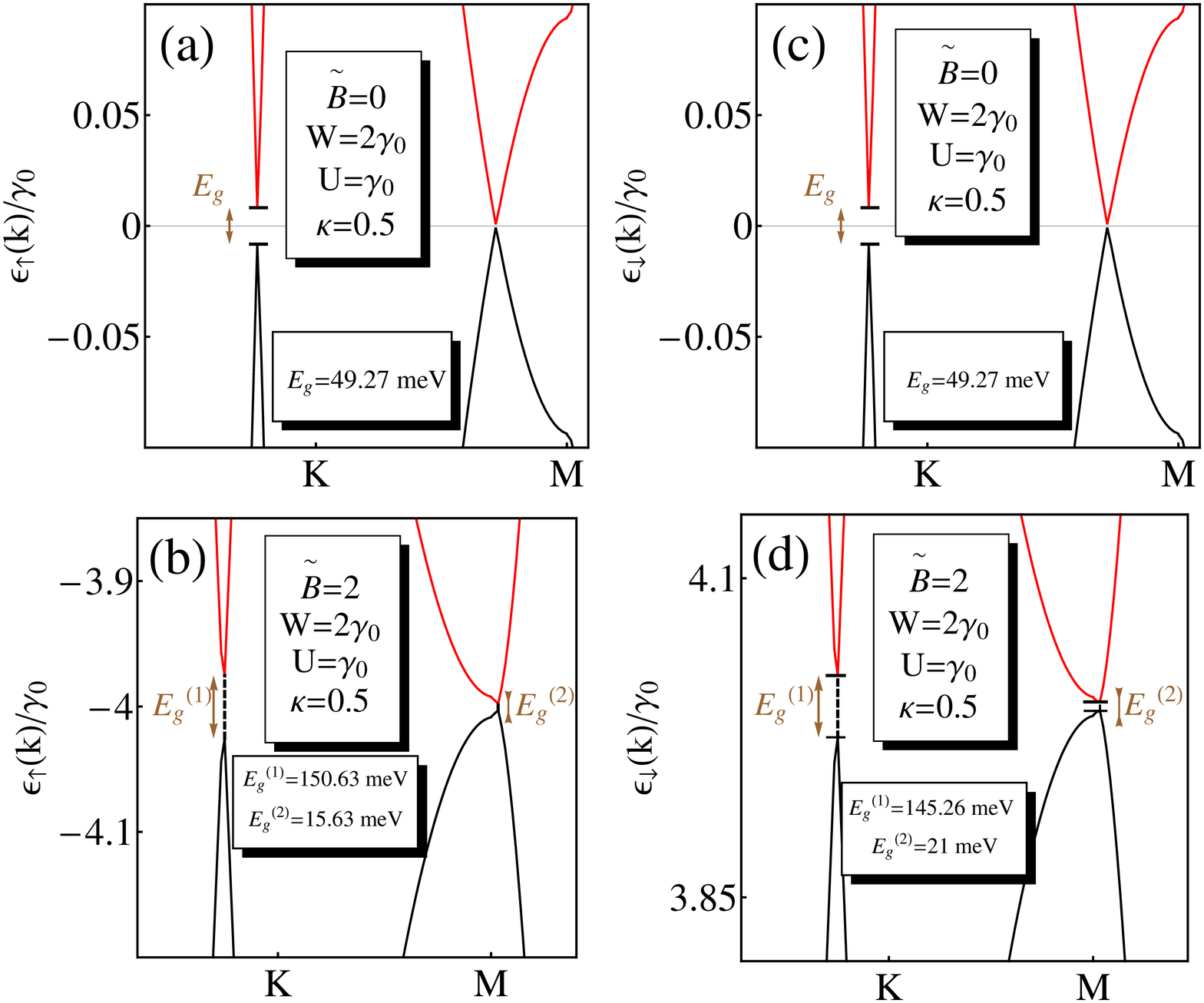}
		\caption{\label{fig:Fig_13}(Color online) Electronic band structure in the AA bilayer graphene, for different values of the interlayer Coulomb interaction parameter $W$ (see in panels (a)-(f)). 
			The low-energy picture was shown here, corresponding to the band structure in Fig.~\ref{fig:Fig_10}. we have shown the principal regions, where the band-gap is opening in the AA-bilayer, caused by the effect of the interlayer Coulomb interaction. The magnetic field is fixed at the value $\tilde{B}=\mu_{B}B/\gamma_0=1$. Four different energy bands are shown in pictures, for each spin direction. The results in top panels (a), (b) and (c) correspond to $\sigma=\uparrow$ and the results in the right panels (d), (e) and (f) correspond to $\sigma=\downarrow$. The half-filling case is considered with $\kappa=0.5$, and the temperature is set at $T=0$.}
	\end{center}
\end{figure} 
%
We observe in Fig.~\ref{fig:Fig_12} that for the partial filling considered there (with $\kappa=1$) the semiconducting state is well defined in the system, for all values of the magnetic field parameter $\tilde{B}$.

Next, we compare the results for the band-gaps, obtained in Fig.~\ref{fig:Fig_12}, with the results of calculations at the half-filling limit, i.e., when $\kappa=0.5$ presented in Fig.~\ref{fig:Fig_13}. As we can see in panels (a)-(c), in Fig.~\ref{fig:Fig_13}, the values of the gaps $E^{(1)}_{g}$ are considerably smaller, in the case of zero magnetic field (we get  $E^{(1)}_{g}=49.27$ meV for both spin channels), and the energy bang-gaps $E^{(2)}_{g}$ are completely absent (i.e., $E^{(2)}_{g}=0$) for $\kappa=0.5$ and $\tilde{B}=0$.

The large values for $E^{(1)}_{g}\left(\sigma\right)$ have been obtained in the limit of high magnetic fields with $\tilde{B}=2$ (see in panels (b) and (d), in Fig.~\ref{fig:Fig_13}) with $E^{(1)}_{g}\left(\sigma=\uparrow\right)=150.63$ meV and $E^{(1)}_{g}\left(\sigma=\downarrow\right)=145.26$ meV. Contrary to the previous case of partial-filling ($\kappa=1$), we have a large decrease of the gaps $E^{(1)}_{g}\left(\sigma=\uparrow\right)$ and $E^{(2)}_{g}\left(\sigma=\downarrow\right)$ for all values of the external magnetic $\tilde{B}$. Thus, for the half-filling case (with $\kappa=0.5$), the increase of the magnetic field parameter leads to the passage from the weak-metallic state into the semiconducting one.   
%
\subsection{\label{sec:Section_4_5} Discussion and perspectives}
%
The results obtained for the exctionic order parameters for different spin directions above show that the AA-BLG system could be an ideal candidate for the observation of the excitonic condensation phenomena at $T=0$. Particularly, in Fig.~\ref{fig:Fig_9}, the excitonic condensate peak was observed in the centrum of the ${\bf{k}}$-space and for the large value of the magnetic fiel parameter.
The numerical results presented in Figs.~\ref{fig:Fig_10}-~\ref{fig:Fig_13} show the new large applicabilities of the biased AA-bilayer graphene systems when it is exposed to the action of the external transverse magnetic field $B$ (see in Fig.~\ref{fig:Fig_1}). Based on the appropriate parameter regime, discussed in the previous sections, the AA-bilayer graphene could be a purposeful candidate for the applications in the modern nanotechnology and solid-state electronics, as a direct band-gap semiconductor with a fully tunable band structure and band-gaps. Both, the Coulomb interaction (which could be tuned by varying the interlayer separation distance \cite{cite_29}) and the magnetic field tunability could be used to obtain the large energy-gaps in the energy spectrum of the AA-BLG. Choosing the parameters to be tuned in the system (for example $W$ or the magnetic field $\tilde{B}$) we can open or close the band-gaps corresponding to $\sigma=\uparrow$ or $\sigma=\downarrow$ which can result in excitations states with only one spin direction $\sigma=\uparrow$ or $\sigma=\downarrow$. This type of situation is achieved in Fig.~\ref{fig:Fig_11} (see in panels (a)-(d)) when by increasing the interaction parameter $W$ and by fixing the magnetic field (namely at the value $\tilde{B}=1$), we close the energy band-gap for the spin direction $\sigma=\uparrow$ (see in panels (a) and (b), in Fig.~\ref{fig:Fig_11}), while the other energy-gap, corresponding to $\sigma=\downarrow$ still very large for all values of the parameter $W$ (see in panels (c) and (d), in the picture), comparable to those in direct band-gap semiconductors \cite{cite_48}. Thus, for the spin direction $\sigma=\uparrow$ we get, in Fig.~\ref{fig:Fig_11}, an inverse transition from semiconducting state into the metallic one (see in panels (a), (b) (c), in Fig.~\ref{fig:Fig_11}). Moreover, in Figs.~\ref{fig:Fig_12}-~\ref{fig:Fig_13}, we have found the transitions of type weak metal-semiconductor and metal-semiconductor, when tuning the magnetic field parameter up to high values. In practice, this type of spin-selective operation mode, with the appropriate band-gap excitations, could be achieved by regulating the photon's emitters (with the appropriate range of wavelength selection) at the corresponding modes $\left(|{\bf{k}}|_{\rm ph},\Omega_{\rm ph}\right)$, where the components $|{\bf{k}}|_{\rm ph}$ and $\Omega_{\rm ph}$ determine the crystallographic direction and the corresponding frequency (or energy) of photons emitted by the source. Thus, the AA-BLG could be also applied as a spin-valve device, where, in one spin direction, we get the blockage of the electron transport (see the result in panels (c) and (d), in Fig.~\ref{fig:Fig_11}, for $\sigma=\downarrow$), while in another spin channel $\sigma=\uparrow$ we get the possibilities for the electron excitations into the conduction band and electronic transport (see in panel (b) in Fig.~\ref{fig:Fig_11}, for $\sigma=\uparrow$). Such a spin-controlled transport in AA-BLG demonstrated here, and the spin-valve effects could be purposeful also for building the new quantum algorithms, important for quantum computations and quantum information theory. 

The results presented here have been performed at the zero-temperature limit. We have shown the AA-BLG system is purposeful for obtaining the stable excitonic condensate states and the metal-semiconductor or semiconductor-metal transitions at zero temperature limit. In this context, the room temperature excitonic and metal-semiconducting transitions in AA-BLG will represent the fundamental steps towards the new possibilities for fast and safe electronics and quantum information.              
%
\section{\label{sec:Section_5} Concluding remarks}
%
In the present paper, we considered the effects of the external magnetic and electric fields on the physical properties in the AA-stacked excitonic bilayer graphene. By using the bilayer generalization of the usual Hubbard model, we derived a set of self-consistent equations for the excitonic order parameters, the average charge density difference between the layers and the chemical potential in the system.
Both partial and half-filling regimes have been discussed and the role of the Hubbard interactions has been revealed. We have calculated the magnetic field dependence of those quantities and we have found the critical value of the magnetic field at which the charge neutrality occurs in the system at the half-filling regime. Moreover, we have shown that above that critical value the excitonic order parameter for the spin direction $\sigma=\uparrow$ gets quenched which opens the new possibility for the spin-controlled electronic transport in the AA-BLG structure and its use as the spin-valve device in modern nanotechnology. Furthermore, the behavior of the chemical potential in the partial-filling regime shows explicitly the possibilities for the excitonic condensation of the preformed excitonic pairs in such a system, not yet observed theoretically, or experimentally.  Additionally, the role of the Hubbard interaction potentials on the average charge density imbalance function has been analyzed and the principal differences between the actions of the Hubbard-$U$ and Hubbard-$W$ couplings have been discussed. 

Indeed, we studied the out-of-plane magnetic field, which causes the Zeeman splitting. Moreover, we considered the case of the uniform and static magnetic field. For this reason, we neglected the orbital magnetic field, which is generally included in the form of the Peierl's phases, associated to the electrons.   

We have shown the effects of the external transverse magnetic field on the excitonic pair formation and condensation in the AA stacked bilayer graphene structure. An ideally isolated excitonic condensate peak appears for the large values of the magnetic field and the excitonic pairing regions have been found in the reciprocal space corresponding to both spin directions. We have found that the excitonic order parameters with the spin directions opposite to the magnetic field are in general larger than those corresponding to the spin direction parallel to the magnetic field. We have calculated the electronic band structure for different values of the external magnetic field $B$ and interlayer Coulomb interaction parameter $W$. We showed that both contribute considerably to the energy spectrum in AA-BLG and result in the opening of the large energy band-gaps in the system comparable to the known values of the band-gaps in the usual semiconducting heterojunctions \cite{cite_48}. We observed that the energy band-gaps are large for the spin direction opposite to the external magnetic field and this observation is true for most values of the magnetic field and interlayer Coulomb interaction parameter, considered in the paper. We have demonstrated that a very-large band-gap is opening (with $E_g=200.97$ meV) even for the non-interacting layers (i.e., when $W=0$) when the external magnetic field is present $B\neq 0$. Furthermore, when augmenting the interlayer interaction potential at the same fixed value of the external magnetic field, the band-gap is decreasing for ${\bm{\sigma}} \uparrow\uparrow {\bf{B}}$ (up to the value $E^{(\uparrow)}_{g}=40.8$ meV) and still very large for the spin directions opposite to the magnetic field ${\bm{\sigma}} \downarrow\uparrow {\bf{B}}$ (with $E^{(\downarrow)}_{g}=116.74$ meV). We have shown that even at $B=0$ a very large energy band-gap is opening in the energy spectrum when considering the interlayer Coulomb interaction $W\neq 0$. Particularly, when $W=2\gamma_0=6$ eV we get $E^{(\uparrow)}_{g}=E^{(\downarrow)}_{g}=119.8$ meV, for partial-filling $\kappa=1$ and $E^{(\uparrow)}_{g}=E^{(\downarrow)}_{g}=49.27$ meV, for the half-filling regime with $\kappa=0.5$. Furthermore, these gaps become large when augmenting the magnetic field parameter. For $\tilde{B}=2$, we get 
very large values for the band-gap parameter $E_{g}$, for both limits of the average electron occupation numbers at the atomic lattice sites positions (partial-filling and half-filling). For the partial-filling, we get $E^{(\uparrow)}_{g}=201.18$ meV and $E^{(\downarrow)}_{g}=204.9$ meV. For the half-filling regime we get $E^{(\uparrow)}_{g}=150.63$ meV and $E^{(\downarrow)}_{g}=145.26$ meV. The general observation coming from our calculations is that the energy band-gap is smaller for the half-filling case and $E^{\sigma}_{g}\left(\kappa=0.5\right)<E^{\sigma}_{g}\left(\kappa=1\right)$. Moreover, a second energy band-gap is opening near the value of the reciprocal wave vector $|{\bf{k}}|=M(2\pi/3,0)$, (the point $M$ at which the usual linear bands cross in the ungapped pristine AA-BLG) and we have also discussed the values of this second energy-gap in the paper. The second small band-gap is zero only in the regimes when $\tilde{B}\neq 0$, $W=0$ (for the partial-filling) or $\tilde{B=0}$, $W\neq0$ (for the half-filling). Thus, we have shown the possibility of metal-semiconductor transition in the biased AA-BLG in the presence of the external transverse magnetic field $B$. We have shown that such transition could happen either by changing the interlayer interaction potential (at the partial-filling and with fixed value of the magnetic field) or by changing the applied magnetic field with the fixed value of the interlayer potential (in the case of the half-filling). Additionally, we established that the semiconducting state is much strong in the limit of partial-filling, i.e., when $\kappa=1$. 

The calculations presented here are still valid for a very large interval of temperatures, although the results are not shown here. Therefore, the results obtained within this paper could be useful also for the studies of the excitonic condensate states and could help also the experimenters to find the excitonic condensates regimes, not yet observed experimentally in the graphene-based heterostructures. The results obtained in the present paper concern the artificially obtained AA-bilayer graphene heterostructures (obtained by deposition from single graphene layers), rather than the pristine or epitaxially obtained BLG.

The spin-selective band-gap formation in the AA-BLG structure, observed in the present paper, opens a new and very interesting future for the technological applications of the AA-BLG as the systems with the spin-valve effect, as quantum spin-selectors for the quantum nanoelectronics, as spin-controlled electronic transistors, quantum spin-transport systems for the quantum information applications and as the spin-injector systems. In our strong conviction, the AA-BLG system under the external field conditions, discussed in the present work, could compete with Bernal stacked AB-BLG structures, due to the large values of the energy band-gaps, approaching the band-gap parameters in the usual direct band-gap semiconducting materials \cite{cite_49}. Another interesting observation, resulting from our calculations, is that the AA-BLG system, considered here, could be used as a system in which the perfect excitonic condensate states could be observed experimentally, in the high magnetic field limit. 

\appendix
%
\section{\label{sec:Section_6} Self-consistent equations for AA-BLG}
%
In the present section, we give the details of calculations of the set of self-consistent (SC) equations in the considered AA-BLG system, discussed in Section \ref{sec:Section_3_2}. In the real-space notations, those SC equations read as
\begin{eqnarray}
	&&\bar{n}_{a}+\bar{n}_{\tilde{a}}=\frac{1}{\kappa},
	\nonumber\\
	&&\bar{n}_{\tilde{a}}-\bar{n}_{a}=\frac{\delta{\bar{n}}}{2},
	\nonumber\\
	&&\Delta_{\sigma}=W\left\langle\bar{\tilde{a}}_{\sigma}\left({\bf{r}}\tau\right)a_{\sigma}\left({\bf{r}}\tau\right)\right\rangle.
	\label{Equation_A_1}
\end{eqnarray}
The average fermionic densities $\bar{n}_{a}$, and $\bar{n}_{\tilde{a}}$ are given as $\bar{n}_{a}=\bar{n}_{a\uparrow}+\bar{n}_{a\downarrow}$ and $\bar{n}_{\tilde{a}}=\bar{n}_{\tilde{a}\uparrow}+\bar{n}_{\tilde{a}\downarrow}$. 
Here, we have considered the excitonic order parameter $\Delta_{\sigma}$ as real, i.e., $\Delta_{\sigma}=\bar{\Delta}_{\sigma}$. This corresponds to the case of a homogeneous AA-BLG system, where the pairing between electrons and holes is translationally invariant. 
The first equation, in the system of equations, in Eq.(\ref{Equation_A_1}), defines the dynamical chemical potential, which should be calculated numerically. The second equation, in the system, in Eq.(\ref{Equation_A_1}), is written for the average charge density difference $\delta{\bar{n}}$ (with $\delta{\bar{n}}=\bar{n}_{2}-\bar{n}_{1}$) between the layers in the AA-BLG.

For calculating the fermionic Green's functions, we rewrite the expression of the partition function in terms of the Grassmann-Nambu spinors, introduced in Section \ref{sec:Section_3_1}
\begin{eqnarray}
	{\cal{Z}}=\int\left[{\cal{D}}\bar{\Psi}\right]\left[{\cal{D}}\Psi\right]e^{-{\cal{S}}\left[\bar{\Psi},\Psi\right]},
	\label{Equation_A_2}
\end{eqnarray}
where the action in exponential on the right-hand side in Eq.(\ref{Equation_A_2}) is given in Eq.(\ref{Equation_17}), in the Section \ref{sec:Section_2_1}. Taking into account two spin directions $\sigma=\uparrow, \downarrow$, we can rewrite for ${\cal{Z}}$ the following expression
\begin{eqnarray}
	{\cal{Z}}={\cal{Z}}_{\uparrow}{\cal{Z}}_{\downarrow}.
	\label{Equation_A_3}
\end{eqnarray}
Furthermore, for the considered spin $\sigma$, we write each component in the product, in Eq.(\ref{Equation_A_3}), in their general forms, which include the auxiliary sources $J\left({\bf{k}}\nu_{n}\right)$ and $\bar{J}\left({\bf{k}}\nu_{n}\right)$
\begin{eqnarray}
	&&{\cal{Z}}_{\sigma}=\int\left[{\cal{D}}\bar{\Psi}_{\sigma}\right]\left[{\cal{D}}\Psi_{\sigma}\right]e^{-\frac{1}{2}\sum_{{\bf{k}}\nu_{n}}\bar{\Psi}_{\sigma}\left({\bf{k}}\nu_{n}\right){\cal{G}}^{-1}_{\sigma}\left({\bf{k}}\nu_{n}\right)\Psi_{\sigma}\left({\bf{k}}\nu_{n}\right)}\times 
	\nonumber\\
	&&\times e^{\frac{1}{2}\sum_{{\bf{k}}\nu_{n}}\bar{\Psi}_{\sigma}\left({\bf{k}}\nu_{n}\right)J_{\sigma}\left({\bf{k}}\nu_{n}\right)+\Psi_{\sigma}\left({\bf{k}}\nu_{n}\right)\bar{J}_{\sigma}\left({\bf{k}}\nu_{n}\right)}.
	\label{Equation_A_4}
\end{eqnarray}
The auxiliary sources $J_{\sigma}\left({\bf{k}}\nu_{n}\right)$ and $\bar{J}_{\sigma}\left({\bf{k}}\nu_{n}\right)$ are the Grassmann-Nambu spinors with $4$-components, which are defined in the same way as the spinors $\bar{\Psi}\left({\bf{k}}\nu_{n}\right)$ and ${\Psi}\left({\bf{k}}\nu_{n}\right)$. Then, we effectuate the HS transformation for the $4$-component Grassmann fields $\bar{\Psi}$ and $\Psi$, in the same way as it is done usually for two-component fermionic field \cite{cite_69}. We get for the two components ${\cal{Z}}_{\uparrow}$ and ${\cal{Z}}_{\downarrow}$, in Eq.(\ref{Equation_A_3})
\begin{eqnarray}
	&&{\cal{Z}}_{\uparrow}=\frac{1}{\prod_{{\bf{k}}\nu_{n}}\det{\cal{G}}_{\uparrow}\left({\bf{k}}\nu_{n}\right)}e^{\frac{1}{2}\sum_{{\bf{k}}\nu_{n}}\bar{J}_{\uparrow}\left({\bf{k}}\nu_{n}\right){\cal{G}}_{\uparrow}\left({\bf{k}}\nu_{n}\right)J_{\uparrow}\left({\bf{k}}\nu_{n}\right)}
	\nonumber\\
	&&{\cal{Z}}_{\downarrow}=\frac{1}{\prod_{{\bf{k}}\nu_{n}}\det{\cal{G}}_{\downarrow}\left({\bf{k}}\nu_{n}\right)}e^{\frac{1}{2}\sum_{{\bf{k}}\nu_{n}}\bar{J}_{\downarrow}\left({\bf{k}}\nu_{n}\right){\cal{G}}_{\downarrow}\left({\bf{k}}\nu_{n}\right)J_{\downarrow}\left({\bf{k}}\nu_{n}\right)}.
	\nonumber\\
	\label{Equation_A_5}
\end{eqnarray}
The functions ${\cal{G}}_{\sigma}\left({\bf{k}}\nu_{n}\right)$, in Eq.(\ref{Equation_A_5}), are the Green's functions for our problem.
Furthermore, the set of SC equations, in Eq.(\ref{Equation_A_1}), will be rewritten in the following expanded form
\begin{eqnarray}
	&&\frac{1}{\kappa}=\frac{1}{\beta{N}}\sum_{{\bf{k}}\nu_{n}}\sum_{m=1,3}\sum_{\sigma}\frac{A_{{mm}\sigma}\left({\bf{k}}\nu_{n}\right)}{\det{\cal{G}}^{-1}_{\sigma}\left({\bf{k}}\nu_{n}\right)},
	\nonumber\\
	&&\frac{\delta{\bar{n}}}{2}=\frac{1}{\beta{N}}\sum_{{\bf{k}}\nu_{n}}\sum_{m=1,3}\sum_{\sigma}i^{{m}+1}\frac{A_{mm\sigma}\left({\bf{k}}\nu_{n}\right)}{\det{\cal{G}}^{-1}_{\sigma}\left({\bf{k}}\nu_{n}\right)},
	\nonumber\\
	&&\Delta_{\sigma}=\frac{W}{\beta{N}}\sum_{{\bf{k}}\nu_{n}}\frac{A_{13\sigma}\left({\bf{k}}\nu_{n}\right)}{\det{\cal{G}}^{-1}_{\sigma}\left({\bf{k}}\nu_{n}\right)}.
	\label{Equation_A_6}
\end{eqnarray}
The functions $A_{11\sigma}\left(x\right)$, $A_{33\sigma}\left(x\right)$ and $A_{13\sigma}\left(x\right)$ in the numerators, in the right-hand sides of these equations, are indeed the polynomials of third ($A_{11\sigma}\left(x\right)$ and $A_{33\sigma}\left(x\right)$) and second ($A_{13\sigma}\left(x\right)$) orders. They are given as 
\begin{eqnarray}
	&&A_{11\sigma}\left(x\right)=x^{3}+a_{1\sigma}x^{2}+b_{1\sigma}\left({\bf{k}}\right)x+c_{1\sigma}\left({\bf{k}}\right),
	\nonumber\\
	&&A_{33\sigma}\left(x\right)=x^{3}+a_{2\sigma}x^{2}+b_{2\sigma}\left({\bf{k}}\right)x+c_{2\sigma}\left({\bf{k}}\right),
	\nonumber\\
	&&A_{13\sigma}\left(x\right)=a_{3\sigma}x^{2}+b_{3\sigma}x+c_{3\sigma}\left({\bf{k}}\right),
	\label{Equation_A_7}
\end{eqnarray} 
where $x=-i\nu_{n}$ and the coefficients $a_{i\sigma}, b_{i\sigma}, c_{i\sigma}$ with $i=1,2,3$ are expressed with the help of parameters $x_{i\sigma}$, defined in Eq.(\ref{Equation_28}), in the Section \ref{sec:Section_3_1}). We obtained
\begin{widetext}
	\begin{eqnarray}
		&&a_{1\sigma}=x_{1\sigma}+2x_{2\sigma}+\frac{V}{2},
		\nonumber\\
		&&b_{1\sigma}\left({\bf{k}}\right)=2x_{1\sigma}x_{2\sigma}+x^{2}_{2\sigma}+x_{1\sigma}V-\frac{V^{2}}{4}-{\tilde{\Delta}}^{2}_{\sigma}-|\tilde{\gamma}_{\bf{k}}|^{2},
		\nonumber\\
		&&c_{1\sigma}\left({\bf{k}}\right)=\frac{1}{8}\left[-\left(V+2x_{2\sigma}\right)\left(V^{2}+2x_{2\sigma}V-2x_{1\sigma}\left(2x_{2\sigma}+V\right)+4{\tilde{\Delta}}^{2}_{\sigma}\right)+4\left(V-2x_{1\sigma}\right)|\tilde{\gamma}_{\bf{k}}|^{2}\right],
		\nonumber\\
		&&a_{2\sigma}=x_{2\sigma}+2x_{1\sigma}-\frac{V}{2},
		\nonumber\\
		&&b_{2\sigma}\left({\bf{k}}\right)=2x_{1\sigma}x_{2\sigma}+x^{2}_{1\sigma}-x_{2\sigma}V-\frac{V^{2}}{4}-{\tilde{\Delta}}^{2}_{\sigma}-|\tilde{\gamma}_{\bf{k}}|^{2},
		\nonumber\\
		&&c_{2\sigma}\left({\bf{k}}\right)=\frac{1}{8}\left[\left(-V+2x_{1\sigma}\right)\left(-V^{2}-2x_{2\sigma}V+2x_{1\sigma}\left(2x_{2\sigma}+V\right)-4{\tilde{\Delta}}^{2}_{\sigma}\right)-4\left(V+2x_{2\sigma}\right)|\tilde{\gamma}_{\bf{k}}|^{2}\right],
		\nonumber\\
		&&a_{3\sigma}={\tilde{\Delta}}_{\sigma},
		\nonumber\\
		&&b_{3\sigma}=\left(x_{1\sigma}+x_{2\sigma}\right){\tilde{\Delta}}_{\sigma},
		\nonumber\\
		&&c_{3\sigma}\left({\bf{k}}\right)=\frac{{\tilde{\Delta}}_{\sigma}}{4}\left[\left(2x_{1\sigma}-V\right)\left(2x_{2\sigma}+V\right)+4|\tilde{\gamma}_{\bf{k}}|^{2}\right].
		\label{Equation_A_8}
	\end{eqnarray}
\end{widetext} 
Next, after summing over the fermionic Matsubara frequencies $\nu_{n}$ in Eq.(\ref{Equation_A_6}), we get the system of self-consistent equations, in Eq.(\ref{Equation_37}), in section \ref{sec:Section_3_2}.  
The coefficients $\alpha_{i{\bf{k}}\sigma}$, $\beta_{i{\bf{k}}\sigma}$ and $\gamma_{i{\bf{k}}\sigma}$ (with $i=1,...4$), in Eq.(\ref{Equation_37}), are defined with the help of the polynomials in Eq.(\ref{Equation_A_7}). We obtain
\begin{widetext}
	\begin{eqnarray}
		\footnotesize
		\arraycolsep=0pt
		\medmuskip = 0mu
		\alpha_{i{\bf{k}}\sigma}
		=(-1)^{i+1}
		\left\{
		\begin{array}{cc}
			\displaystyle  & \frac{A_{11\sigma}\left(\epsilon_{i\sigma}\left({\bf{k}}\right)\right)}{\epsilon_{1\sigma}\left({\bf{k}}\right)-\epsilon_{2\sigma}\left({\bf{k}}\right)}\prod_{j=3,4}\frac{1}{\epsilon_{i\sigma}\left({\bf{k}}\right)-\epsilon_{j\sigma}\left({\bf{k}}\right)},  \ \ \  $if$ \ \ \ i=1,2,
			\newline\\
			\newline\\
			\displaystyle  & \frac{A_{11\sigma}\left(\epsilon_{i\sigma}\left({\bf{k}}\right)\right)}{\epsilon_{3\sigma}\left({\bf{k}}\right)-\epsilon_{4\sigma}\left({\bf{k}}\right)}\prod^{}_{j=1,2}\frac{1}{\epsilon_{i\sigma}\left({\bf{k}}\right)-\epsilon_{j\sigma}\left({\bf{k}}\right)},  \ \ \  $if$ \ \ \ i=3,4,
		\end{array}\right.
		\label{Equation_A_9}
	\end{eqnarray}
	\begin{eqnarray}
		\footnotesize
		\arraycolsep=0pt
		\medmuskip = 0mu
		\beta_{i{\bf{k}}\sigma}
		=(-1)^{i+1}
		\left\{
		\begin{array}{cc}
			\displaystyle  & \frac{A_{33\sigma}\left(\epsilon_{i\sigma}\left({\bf{k}}\right)\right)}{\epsilon_{1\sigma}\left({\bf{k}}\right)-\epsilon_{2\sigma}\left({\bf{k}}\right)}\prod_{j=3,4}\frac{1}{\epsilon_{i\sigma}\left({\bf{k}}\right)-\epsilon_{j\sigma}\left({\bf{k}}\right)},  \ \ \  $if$ \ \ \ i=1,2,
			\newline\\
			\newline\\
			\displaystyle  & \frac{A_{33\sigma}\left(\epsilon_{i\sigma}\left({\bf{k}}\right)\right)}{\epsilon_{3\sigma}\left({\bf{k}}\right)-\epsilon_{4\sigma}\left({\bf{k}}\right)}\prod^{}_{j=1,2}\frac{1}{\epsilon_{i\sigma}\left({\bf{k}}\right)-\epsilon_{j\sigma}\left({\bf{k}}\right)},  \ \ \  $if$ \ \ \ i=3,4,
		\end{array}\right.
		\label{Equation_A_10}
	\end{eqnarray}
	and 
	\begin{eqnarray}
		\footnotesize
		\arraycolsep=0pt
		\medmuskip = 0mu
		\gamma_{i{\bf{k}}\sigma}
		=(-1)^{i+1}
		\left\{
		\begin{array}{cc}
			\displaystyle  & \frac{A_{13\sigma}\left(\epsilon_{i\sigma}\left({\bf{k}}\right)\right)}{\epsilon_{1\sigma}\left({\bf{k}}\right)-\epsilon_{2\sigma}\left({\bf{k}}\right)}\prod_{j=3,4}\frac{1}{\epsilon_{i\sigma}\left({\bf{k}}\right)-\epsilon_{j\sigma}\left({\bf{k}}\right)},  \ \ \  $if$ \ \ \ i=1,2,
			\newline\\
			\newline\\
			\displaystyle  & \frac{A_{13\sigma}\left(\epsilon_{i\sigma}\left({\bf{k}}\right)\right)}{\epsilon_{3\sigma}\left({\bf{k}}\right)-\epsilon_{4\sigma}\left({\bf{k}}\right)}\prod^{}_{j=1,2}\frac{1}{\epsilon_{i\sigma}\left({\bf{k}}\right)-\epsilon_{j\sigma}\left({\bf{k}}\right)},  \ \ \  $if$ \ \ \ i=3,4.
		\end{array}\right.
		\label{Equation_A_11}
	\end{eqnarray}
\end{widetext}

By solving numerically the system of equations in Eq.(\ref{Equation_A_6}) we get the important physical parameters in the system, such as the energy necessary for the single-quasiparticle excitations (creation or annihilation) $\mu$, the average charge density imbalance function $\delta{\bar{n}}$, which describes the dynamical changes of the average electron densities in the layers, and the excitonic order parameter $\Delta_{\sigma}$, as well.  
%
\section{\label{sec:Section_7} Hubbard-Stratanovich decoupling of the non-linear density terms}
%
\subsection{\label{sec:Section_7_1} Decoupling of the Coulomb-$U$ term}
%
Now we effectuate the Hubbard-Stratanovich (HS) transformation of the non-linear density terms in the Hamiltonian, in Eq.(\ref{Equation_1}).
The density operators in Eqs.(\ref{Equation_6}),(\ref{Equation_8}) and (\ref{Equation_9}) are now given in terms of simple Grassmann complex variables, and we can write for the product $n_{\eta\uparrow}n_{\eta\downarrow}$ in the $U$-term in Eq.(\ref{Equation_4}) the following relation 
\begin{eqnarray}
	n_{\eta\uparrow}({\bf{r}}\tau)n_{\eta\downarrow}({\bf{r}}\tau)=\frac{1}{4}\left[n^{2}_{\eta}({\bf{r}}\tau)-\left({{\bf{n}}}_{\eta}\left({\bf{r}}\tau\right)\hat{\sigma}_{\rm z}\right)^{2}\right].
	\label{Equation_B_1}
\end{eqnarray}
The vector-operator $\hat{{\bf{n}}}_{\eta}\left({\bf{r}}\right)$, in Eq.(\ref{Equation_B_1}), is a two-dimensional vector, given as
\begin{eqnarray}
	\hat{{\bf{n}}}_{\eta}\left({\bf{r}}\right)=\left(\hat{n}_{\eta\uparrow}\left({\bf{r}}\right),\hat{n}_{\eta\downarrow}\left({\bf{r}}\right)\right).
	\label{Equation_B_2}
\end{eqnarray}	
The product ${\bf{n}}_{\eta}\left({\bf{r}}\tau\right)\hat{\sigma}_{\rm z}$, in the right-hand side in Eq.(\ref{Equation_B_1}), is indeed the charge density difference between different spin configurations $\sigma=\uparrow$ and $\sigma=\downarrow$, i.e.,  
\begin{eqnarray}
	{\bf{n}}_{\eta}\left({\bf{r}}\tau\right)\hat{\sigma}_{\rm z}= {n}_{\eta\uparrow}\left({\bf{r}}\tau\right)-{n}_{\eta\downarrow}\left({\bf{r}}\tau\right).
	\label{Equation_B_3}
\end{eqnarray}
Then, the HS transformation looks like 
\begin{eqnarray}
	&&e^{-\frac{U}{4}\sum_{{\bf{r}}}\int^{\beta}_{0}d\tau n^{2}_{\eta}\left({\bf{r}}\tau\right)}
	\nonumber\\
	&&=\int\left[{\cal{D}}V_{\eta}\right]e^{-\sum_{{\bf{r}}}\int^{\beta}_{0}d\tau\left(\frac{V^{2}_{\eta}\left({\bf{r}}\tau\right)}{U}-in_{\eta}\left({\bf{r}}\tau\right)V_{\eta}\left({\bf{r}}\tau\right)\right)}.
	\label{Equation_B_4}
\end{eqnarray}
The integration in right-hand side in Eq.(\ref{Equation_B_4}) is evaluated over the auxiliary complex field $V_{\eta}\left({\bf{r}}\tau\right)$. We can calculate this integral by using the saddle-point approximation of it. Indeed this procedure is equivalent, in some sense, to the usual mean-field approximation. Moreover, the procedure described here is more precise, since it deals with a more general form of the generating function. In the exponential, on the right-hand side in Eq.(\ref{Equation_B_4}), we have a function $f\left[V_{\eta},n_{\eta}\right]$, given as
\begin{eqnarray}
	f\left[V_{\eta},n_{\eta}\right]=\sum_{{\bf{r}}}\int^{\beta}_{0}d\tau\left(-\frac{V^{2}_{\eta}\left({\bf{r}}\tau\right)}{U}+in_{\eta}\left({\bf{r}}\tau\right)V_{\eta}\left({\bf{r}}\tau\right)\right)
	\nonumber\\
	\label{Equation_B_5}
\end{eqnarray}
and we approximate the integral as 
\begin{eqnarray}
	\int\left[{\cal{D}}V_{\eta}\right]e^{f\left[V_{\eta},n_{\eta}\right]}=e^{f\left[V_{0\eta},n_{\eta}\right]},
	\label{Equation_B_6}
\end{eqnarray}
where $V_{0\eta}$ is the saddle-point value of the variable $V_{\eta}\left({\bf{r}}\tau\right)$. We calculate the saddle-point value by solving the equation
\begin{eqnarray}
	f'\left[V_{\eta}\right]=0.
	\label{Equation_B_7}
\end{eqnarray} 
Then we get for $V_{0\eta}$ 
\begin{eqnarray}
	V_{0\eta}=\frac{iU}{2}\left\langle{n}_{\eta}\right\rangle.
	\label{Equation_B_8}
\end{eqnarray} 
Here, $\left\langle{n}_{\eta}\right\rangle$ is the grand canonical average of the density function ${n}_{\eta}\left({\bf{r}}\tau\right)$. It could be calculated exactly with the help of partition function, in Eq.(\ref{Equation_12}), as 
\begin{eqnarray}
	\left\langle ... \right\rangle=\frac{1}{{\cal{Z}}}\int\prod_{\eta}\left[{\cal D}\bar{\eta}{\cal{D}}\eta\right]...e^{-{\cal{S}}}.
	\label{Equation_B_9}
\end{eqnarray} 
Next, the contribution to the action in Eq.(\ref{Equation_13}), coming from the decoupling, given in Eq.(\ref{Equation_B_4}) is
\begin{eqnarray}
	{\cal{S}}\left[V_{0\eta}\right]=-\sum_{{\bf{r}}}\int^{\beta}_{0}d\tau\frac{U}{2}\bar{n}_{\eta}n_{\eta}\left({\bf{r}}\tau\right)
	\label{Equation_B_10}
\end{eqnarray}
and the contribution to total Hamiltonian in Eq.(\ref{Equation_1}) is 
\begin{eqnarray}
	\Delta{{\cal{H}}}_{U}=\frac{U}{2}\sum_{{\bf{r}}}\bar{n}_{\eta}n_{\eta}\left({\bf{r}}\tau\right).
	\label{Equation_B_11}
\end{eqnarray} 
The decoupling of quadratic charge density difference term $(U/4)\int^{\beta}_{0}d\tau\sum_{{\bf{r}}}\left(\hat{{\bf{n}}}_{\eta}\left({\bf{r}}\tau\right)\hat{\sigma}_{\rm z}\right)^{2}$ is also obvious
\begin{eqnarray}
	&&e^{\frac{U}{4}\sum_{{\bf{r}}}\int^{\beta}_{0}d\tau\left({{\bf{n}}}_{\eta}\left({\bf{r}}\tau\right)\hat{\sigma}_{\rm z}\right)^{2}}
	\nonumber\\
	&&=\int\left[{\cal{D}}\zeta\right]e^{-\sum_{{\bf{r}}}\int^{\beta}_{0}d\tau\left(\frac{\zeta^{2}_{\eta}\left({\bf{r}}\tau\right)}{U}-\zeta_{\eta}\left({\bf{r}}\tau\right){{\bf{n}}}_{\eta}\left({\bf{r}}\tau\right)\hat{\sigma}_{\rm z}\right)}.
	\label{Equation_B_12}
\end{eqnarray} 
Within the same saddle-point approximation procedure, we can calculate the average values of the decoupling field $\zeta_{\eta}\left({\bf{r}}\tau\right)$ (Indeed, the average value, obtained within this method, is equivalent to the one calculated within the usual mean-field theory). We get
\begin{eqnarray}
	\zeta_{0\eta}=\frac{U}{2}\left\langle{{\bf{n}}_{\eta}\left({\bf{r}}\tau\right)\hat{\sigma}_{\rm z}}\right\rangle.
	\label{Equation_B_13}
\end{eqnarray} 
Indeed, we suppose the equal average spin populations in the system at equilibrium, thus, we have $\left\langle {n}_{\eta \uparrow}\left({\bf{r}}\tau\right)\right\rangle=\left\langle {n}_{\eta \downarrow}\left({\bf{r}}\tau\right)\right\rangle$, therefore, we get $\left\langle{{\bf{n}}_{\eta}\left({\bf{r}}\tau\right)\hat{\sigma}_{\rm z}}\right\rangle=0$. Thus, we put $\zeta_{0\eta}=0$. 
%
\subsection{\label{sec:Section_7_2} Decoupling of the Coulomb-$W$ term}
%
A simple operator calculus shows that the interlayer Coulomb interaction term-$W$ can be rewritten in the more efficient form by using the auxiliary excitonic operators 
\begin{eqnarray}
	\hat{\xi}^{(a)}_{\sigma\sigma'}\left({\bf{r}}\right)=\hat{a}^{\dag}_{\sigma}\left({\bf{r}}\right)\hat{\tilde{a}}_{\sigma'}\left({\bf{r}}\right)
	\label{Equation_B_14}
\end{eqnarray}
and
\begin{eqnarray}
	\hat{\xi}^{(b)}_{\sigma\sigma'}\left({\bf{r}}\right)=\hat{{{b}}}^{\dag}_{\sigma'}\left({\bf{r}}\right)\hat{\tilde{b}}^{\dag}_{\sigma}\left({\bf{r}}\right).
	\label{Equation_B_15} 
\end{eqnarray}
We can write
\begin{eqnarray}
	{\cal{H}}_{\rm W}=2W\sum_{{\bf{r}}}\sum_{\eta=a,b}\hat{n}_{\eta}\left({\bf{r}}\right)-W\sum_{{\bf{r}}\sigma\sigma'\lambda}|\hat{\xi}^{(\lambda)}_{\sigma\sigma'}\left({\bf{r}}\right)|^{2},
	\nonumber\\
	\label{Equation_B_16}
\end{eqnarray} 
where $\lambda=a,b$, according to definitions in Eqs.(\ref{Equation_B_14}) and (\ref{Equation_B_15}).
Then we can decouple the biquadratic fermionic term in Eq.(\ref{Equation_B_16}) by employing again the Grassmann-field path integration techniques and then the saddle-point approximation for the auxiliary fermionic fields $\bar{\Delta}^{(\lambda)}_{\sigma\sigma'}\left({\bf{r}}\tau\right)$ and ${\Delta}^{(\lambda)}_{\sigma\sigma'}\left({\bf{r}}\tau\right)$, introduced above. The HS decoupling of the $W$-term is 
\begin{eqnarray}
	&&\exp\left(W\sum_{{\bf{r}}}\sum_{\sigma\sigma'}\sum_{\lambda}\int^{\beta}_{0}d\tau|{\xi}^{(\lambda)}_{\sigma\sigma'}\left({\bf{r}}\tau\right)|^{2}\right)=
	\nonumber\\
	&&=\int\prod_{\lambda}\left[{\cal{D}}\bar{\Delta}^{(\lambda)}{\cal{D}}{\Delta}^{(\lambda)}\right]e^{-\frac{1}{W}\sum_{{\bf{r}}}\sum_{\sigma\sigma'}\sum_{\lambda}\int^{\beta}_{0}d\tau|{\Delta}^{(\lambda)}_{\sigma\sigma'}\left({\bf{r}}\tau\right)|^{2}}\times
	\nonumber\\	
	\times &&e^{\sum_{{\bf{r}}}\sum_{\sigma\sigma'}\sum_{\lambda}\int^{\beta}_{0}d\tau\left(\bar{\Delta}^{(\lambda)}_{\sigma\sigma'}\left({\bf{r}}\tau\right){{\xi}}^{(\lambda)}_{\sigma\sigma'}\left({\bf{r}}\tau\right)+{\Delta}^{(\lambda)}_{\sigma\sigma'}\left({\bf{r}}\tau\right)\bar{{\xi}}^{(\lambda)}_{\sigma\sigma'}\left({\bf{r}}\tau\right)\right)}.
	\nonumber\\
	\label{Equation_B_17}
\end{eqnarray} 
Next, we consider the function in exponential on the right-hand side in Eq.(\ref{Equation_B_17}) and we replace the integral by the saddle-point value of the exponential function at the points $\bar{{\Delta}}^{(\lambda)}_{0\sigma\sigma'}$ and ${{\Delta}}^{(\lambda)}_{0\sigma\sigma'}$. In turn, those saddle-point values could be obtained after functional derivation of the integral in the right-hand side in Eq.(\ref{Equation_B_17}), and we get the following values
\begin{eqnarray}
	{{\Delta}}^{(\lambda)}_{0\sigma\sigma'}=W\left\langle{{\xi}}^{(\lambda)}_{\sigma\sigma'}\left({\bf{r}}\tau\right)\right\rangle,
	\nonumber\\	
	\bar{{\Delta}}^{(\lambda)}_{0\sigma\sigma'}=W\left\langle\bar{{\xi}}^{(\lambda)}_{\eta\sigma\sigma'}\left({\bf{r}}\tau\right)\right\rangle.
	\label{Equation_B_18}
\end{eqnarray} 
Indeed, the parameters ${{\Delta}}^{(\lambda)}_{0\sigma\sigma'}$ and $\bar{{\Delta}}^{(\lambda)}_{0\sigma\sigma'}$, obtained above, represent the mean-field values of the excitonic gap parameters $\Delta^{(\lambda)}_{\sigma\sigma'}$ and $\bar{\Delta}^{(\lambda)}_{\sigma\sigma'}$, i.e.,
\begin{eqnarray}
	{{\Delta}}^{(\lambda)}_{0\sigma\sigma'}=\Delta^{(\lambda)}_{\sigma\sigma'}
	\nonumber\\	
	\bar{{\Delta}}^{(\lambda)}_{0\sigma\sigma'}=\bar{\Delta}^{(\lambda)}_{\sigma\sigma'}.
	\label{Equation_B_19}
\end{eqnarray}
Furthermore, we suppose them as real and we put $\sigma=\sigma'$ (this corresponds to the case of pairing between electrons and holes with opposite spin directions). On the other hand, for the homogeneous AA-BLG system, we have ${\Delta}^{(a)}_{\sigma\sigma'}={\Delta}^{(b)}_{\sigma\sigma'}$ and we can omit the sublattice index $\lambda$. Thus, we have
\begin{eqnarray}
	\bar{\Delta}^{(\lambda)}_{\sigma\sigma'}={\Delta}^{(\lambda)}_{\sigma\sigma'}={\Delta}^{(\lambda)}_{\sigma} \equiv \Delta_{\sigma}.
	\label{Equation_B_20}
\end{eqnarray}
Then, the contribution to total Hamiltonian in Eq.(\ref{Equation_1}) coming from the decoupling procedure will be 
\begin{eqnarray}
	\Delta{{\cal{H}}}_{W}=-\sum_{{\bf{r}}\sigma}\sum_{\lambda}{\Delta}_{\sigma}{\xi}^{(\lambda)}_{\sigma}\left({\bf{r}}\right).
	\label{Equation_B_21}
\end{eqnarray} 
It is interesting to remark, at the end of this Section, that the quadratic terms in $\left\langle{n}_{\eta}\right\rangle$, $\Delta_{0\eta\sigma\sigma'}$ and $\bar{\Delta}_{0\eta\sigma\sigma'}$, appearing when putting back those saddle-point values in the respective functions in exponentials, give just the constant contributions to the total Hamiltonian of the system and, therefore, we neglect them, for the first treatment. 
%

\end{document}